\documentclass[final,5p,times,twocolumn]{elsarticle}

\usepackage{url}
\usepackage{draftcopy}

\journal{Astroparticle Physics}

\begin{document}

\begin{frontmatter}



\title{Surveys with the Cherenkov Telescope Array}


\author[grenoble]{G. Dubus}
\author[m]{J. L. Contreras}
\author[kipac]{S. Funk}
\author[lupm]{Y. Gallant}
\author[m]{T. Hassan}
\author[leicester]{J. Hinton}
\author[kyoto,kipac]{Y. Inoue}
\author[toulouse]{J. Kn\"odlseder}
\author[grenoble]{P. Martin}
\author[m]{N. Mirabal}
\author[mdn]{M. de Naurois}
\author[lupm]{M. Renaud}
\author{for the CTA Consortium}
\address[grenoble]{UJF-Grenoble 1 / CNRS-INSU, Institut de Plan\'etologie et d'Astrophysique de Grenoble (IPAG) UMR 5274, Grenoble, F-38041, France}
\address[m]{Dpto. de F\'isica At\'omica, Molecular y Nuclear, Universidad Complutense de Madrid, Spain}
\address[kipac]{W. W. Hansen Experimental Physics Laboratory, KIPAC, Department of Physics and SLAC, Stanford University, Stanford, CA 94305, USA}
\address[lupm]{Universit\'e Montpellier 2 / CNRS-IN2P3, Laboratoire Univers et Particules de Montpellier (LUPM), UMR 5207, Montpellier, F-34095, France}
\address[leicester]{Department of Physics and Astronomy, University of Leicester, Leicester LE1 7RH, UK}
\address[kyoto]{Department of Astronomy, Kyoto University, Kitashirakawa, Sakyo-ku, Kyoto 606-8502, Japan}
\address[toulouse]{Universit\'e de Toulouse / CNRS-INSU, Institut de Recherche en Astrophysique et Plan\'etologie (IRAP) UMR 5277, Toulouse, F-31028, France}
\address[mdn]{Ecole Polytechnique / CNRS-IN2P3,Laboratoire Leprince-Ringuet (LLR), UMR 7638, Palaiseau, F-91128, France}
\begin{abstract}
Surveys open up unbiased discovery space and generate legacy datasets of long-lasting value. One of the goals of imaging arrays of Cherenkov telescopes like CTA is to survey areas of the sky for faint very high energy gamma-ray (VHE) sources, especially sources that would not have drawn attention were it not for their VHE emission ({\em e.g.} the Galactic ``dark accelerators"). More than half the currently known VHE sources are to be found in the Galactic plane. Using standard techniques, CTA can carry out a survey of the region $|\ell|\leq 60^\circ$, $|b|\leq 2^\circ$ in 250 hr (1/4th the available time per year at one location) down to a uniform sensitivity of 3 mCrab (a ``Galactic Plane survey"). CTA could also survey 1/4th of the sky down to a sensitivity of 20 mCrab in 370 hr of observing time (an ``all-sky survey"), which complements well the surveys by the {\em Fermi}/LAT at lower energies and extended air shower arrays at higher energies. Observations in (non-standard) divergent pointing mode may shorten the ``all-sky survey'' time to about 100 hr with no loss in survey sensitivity. We present the scientific rationale for these surveys, their place in the multi-wavelength context, their possible impact and their feasibility. We find that the Galactic Plane survey has the potential to detect hundreds of sources. Implementing such a survey should be a major goal of CTA. Additionally, about a dozen blazars, or counterparts to {\em Fermi}/LAT sources, are expected to be detected by the all-sky survey, whose prime motivation is the search for extragalactic ``dark accelerators". 
\end{abstract}


\end{frontmatter}


\section{Introduction}
Surveys constitute an unbiased, systematic exploratory approach; they favour discoveries of unknown source classes; they allow for scheduling ease and homogeneous data reduction; they provide legacy datasets for future reference. Surveys of different extents and depths are amongst the scientific goals of all major facilities that are planned or in operation. This is particularly critical for observational domains that are opening up, such as very high energy (VHE $\geq$30 GeV) gamma rays, with wide scope for surprises. Indeed, the Galactic Plane survey carried out by HESS led to the detection of dozens of sources, many of which were unexpected; among these, the {\em dark accelerators}, have no obvious counterparts at other wavelengths \citep{Aharonian:2005dx,2006ApJ...636..777A,Aharonian:2008ap}. In high energy gamma rays (HE $\geq$30 MeV), the {\em Fermi}/LAT catalog \citep{2010ApJS..188..405A} has a major impact on our knowledge of the HE sky with statistical studies rendered possible for several classes of sources (blazars, pulsars, globular clusters and normal galaxies), with HE emission associated with unexpected objects (e.g. nova V407 Cyg), with $\approx 30\%$ of the 1873 HE sources listed in the second catalog unassociated with known objects \cite{2011arXiv1108.1435T}. 

Compared to previous imaging arrays of Cherenkov telescopes (IACTs), surveys with CTA can only benefit from the increased sensitivity (detection of fainter sources), larger field-of-view (to study multiple or extended sources), improved angular resolution (to alleviate source confusion), broader energy range and better energy resolution (to help determination of the source spectral energy distribution). Surveys provide an immense, if not necessary, service to the research community in the context of an open observatory.  Surveys constitute versatile datasets that enable the detection of unexpected sources and provide testing ground for new theoretical ideas. Surveys are an indispensable tool to assist the community in formulating open time proposals for in-depth studies.

Here, we review current work and perspectives on possible surveys with CTA, their advantages and drawbacks, their relationship with current state-of-the-art and their place in the multi-wavelength context. More precisely, we focus on two easily-defined general purpose surveys that may serve as flagship projects for CTA:  a deep {\em Galactic Plane survey} and a more shallow, wider {\em all-sky survey} (both being limited in practice by the fraction of the sky accessible at zenith angle $\leq 60^\circ$ from the chosen CTA sites in the Northern and Southern hemispheres). The general scientific objectives and multi-wavelength context are described in \S2. Simulations have been carried out to study the implementation and achievable sensitivities of these surveys using the latest response files for CTA (\S3). Their potential in terms of number of detections to expect, based on the current knowledge of various source populations, is then presented in \S4.  We conclude on the strengths and limitations of both survey proposals.

\section{Scope and motivation for CTA surveys}

\begin{figure}
\resizebox{\hsize}{!}{\includegraphics{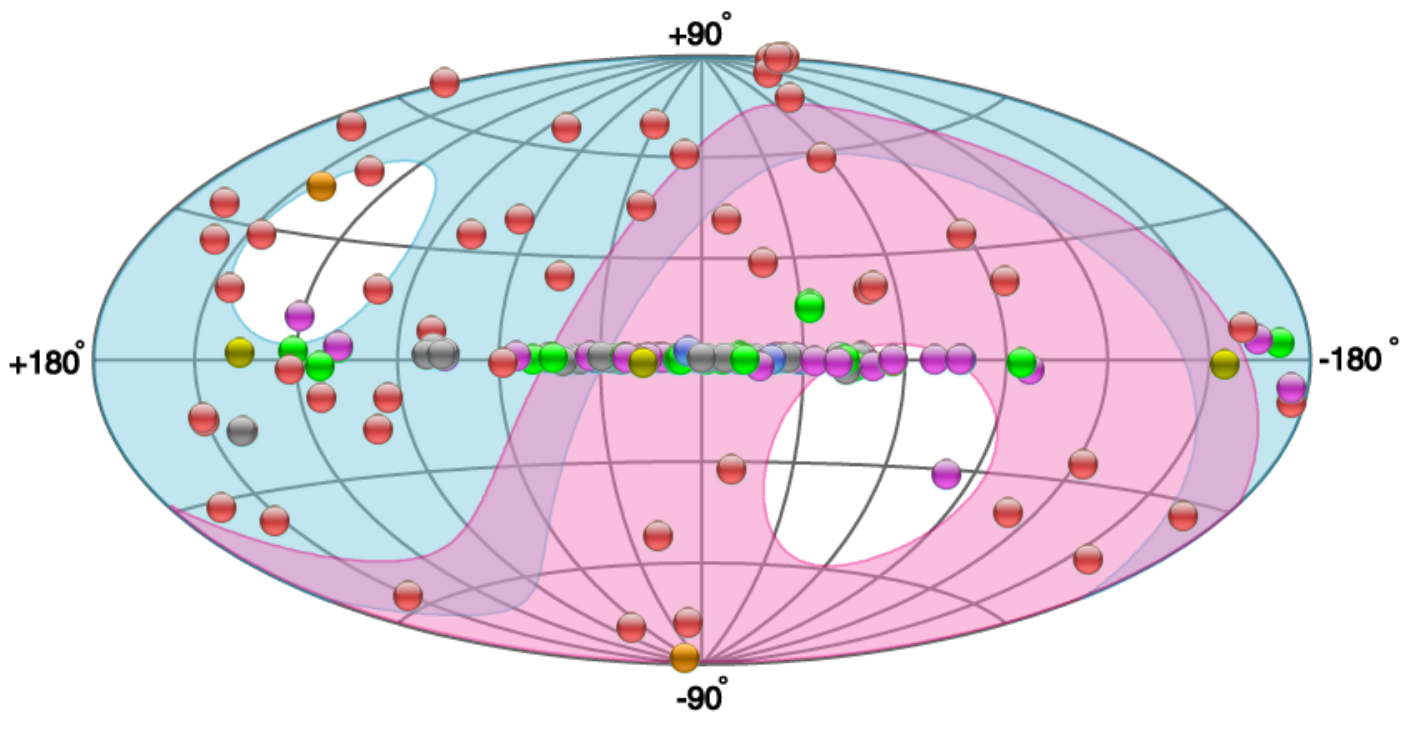}}
\caption{Known VHE sources as of July 2011 from the TeVCat catalog, plotted in Galactic coordinates. The colored regions show the accessible sky from the HESS (red) and Veritas/MAGIC (blue) sites.  Point color identifies sources type: pulsar wind nebulae (magenta), AGN (red), SNR (green), binaries (yellow), starburst (orange), other identified (blue), unidentified (grey). The proposed CTA southern sites (Argentina, Namibia) cover almost the same region of the sky as HESS. The blind spots correspond to zenith angles $>50^\circ$.  \label{tevcat}}
\end{figure}

\subsection{A CTA Galactic Plane Survey\label{gps}}
More than half of the currently known VHE sources  are located within a few degrees of the Galactic Plane: 69 are within $|b|\leq 2^\circ$ out of the 136 VHE sources listed in the TeVCat catalog\footnote{See \url{http://tevcat.uchicago.edu} and\\ \url{http://www.mpp.mpg.de/~rwagner/sources/} (see Fig.~\ref{tevcat})}. The spatial density of VHE sources is greater close to the Galactic Plane (56 sources with $|b|\leq 2^\circ$ and $|\ell|\leq 60^\circ$), even if there is a bias due to the larger exposure accumulated close to the Galactic Plane. VHE Galactic sources are, but for a few exceptions (Galactic Center, gamma-ray binaries, blazars), extended and non-variable making a Galactic Plane survey an attractive solution to maximize coverage and observing efficiency without losing sources.

The first Galactic Plane survey at VHE energies by an imaging array of Cherenkov telescopes (IACT) was carried out by the HESS collaboration \citep{Aharonian:2005dx}. The initial survey used 230 hr of livetime to cover $\left|\ell\right|\leq \pm30^{\circ}$, $\left|{b}\right|\leq 3^{\circ}$ ($\approx$ 0.9\% of the sky). This survey led to the detection of 17 sources (including 3 previously known sources) with the faintest ones having a VHE flux equivalent to $\approx$ 5\% of that of the Crab nebula (50 mCrab). The survey has now been extended to cover $-90^{\circ}<\ell<60^{\circ}$ ($\approx$ 2.2\% of the sky) with the detection of close to 50 sources and a sensitivity reaching 20 mCrab \citep{2009arXiv0907.0768C}. The current Galactic Plane survey is estimated to be complete down to $\approx 85$ mCrab \citep{2009arXiv0905.1287R}. The total observing time reaches 1500 hr {\em i.e.} about 1.5 years of available observing time. The {\em Milagro} collaboration surveyed the region $30^{\circ}<\ell<120^{\circ}$, $|{b}|\leq 10^{\circ}$ ($\approx$ 7.3\% of the sky) using 2300 days $\approx 6.3$ years of observations with a water Cherenkov extensive air shower (EAS) array. In total, {\em Milagro} detected 8 sources above a median energy of 20 TeV, down to a sensitivity $\approx 200$ mCrab, some of which have an extension of several degrees \citep{2007ApJ...664L..91A}.

Galactic Plane surveys are well suited to IACTs given the limited area to cover, as well as their lower energy thresholds and lower confusion levels compared to EAS arrays. A ten-fold improvement in sensitivity means that CTA, with an investment in time similar to the initial HESS survey (250 hr), can reach at least 5 mCrab over a similar sized region of the Galactic Plane. Detailed simulations show that a 3 mCrab sensitivity is achieved (\S3). Such a sensitivity is equivalent to the deepest $\approx$100 -- 200 hr exposures that are carried out by the current generation of IACTs on a few selected objects ({\em e.g.} the supernova remnant SN\,1006).  More than 300 sources are expected at a sensitivity of 2 mCrab based on an extrapolation of the current $\log N- \log S$ diagram for VHE Galactic sources \citep{2009arXiv0905.1287R} (see \S\ref{mr} below). Half of the sources in the TeVCat catalog are within 1.5$^{\circ}$ of the Galactic Plane. Few sources have been detected further away from the Plane by the HESS survey: one exception is  HESS J1507-622 at $b=-3.5^{\circ}$ \citep{2011A&A...525A..45H}. The density of known VHE sources also increases closer to the Galactic Center, favoring the use of the CTA array in the southern site for such an exploration (Fig.~\ref{tevcat}).  Both proposed sites for the southern CTA array cover well the central regions of our galaxy. However, a full exploration of the Galactic Plane requires both southern and northern array.

A CTA Galactic Plane survey would give access to dozens of supernova remnants (SNRs) and pulsar wind nebulae (PWNe) with no pointing {\em a priori} \citep{2008APh....29...63C,2008AIPC.1085..886F}, enabling the first population studies of these objects at VHE ({\em e.g.} L$_\mathrm{VHE}$ vs SNR age or vs pulsar power). Such a dataset can be used to search for emission from cosmic ray interaction with molecular clouds, stellar clusters, dark accelerators or binaries (with the caveat that the latter are variable). Once the most promising sources have been identified, dedicated pointed observations can be requested for detailed spectro-imaging or variability monitoring \citep{cta_pwn, cta_snr, cta_binaries}.

\subsection{A CTA all-sky survey\label{allsky}}
All-sky VHE surveys are well suited to water Cherenkov extensive air shower (EAS) arrays that observe the whole sky with high duty cycles.  Large surveys with IACTs are hampered by the low observation duty cycle (night time and moonlight constraints) and limited field-of-views (few degrees). The {\em Milagro} and Tibet air shower arrays have carried out a survey for sources in the Northern hemisphere down to an average sensitivity of 600 mCrab above 1 TeV \cite{2004ApJ...608..680A,2005ApJ...633.1005A}. The HAWC project aims for a sensitivity to 1 Crab sources in a day (50 mCrab in a year), a median energy around a TeV and a $1^{\circ}$ angular resolution\footnote{HAWC website at \url{http://hawc.umd.edu/}}. EAS arrays have lower angular resolution ($\approx 1^{\circ}$), higher energy thresholds ($\geq$ 1 TeV) and are less sensitive than IACTs. Yet, because of their high duty cycles and large field-of-views, EAS arrays remain irreplaceable tools to study the transient VHE sky. Still, with an increased sensitivity and larger field-of-view compared to the current generation of IACTs, a large-scale CTA survey would bring improvements in survey depth, energy threshold and angular resolution over the HAWC map even with a moderate investment in time. 

For an IACT, a quarter of the sky (10$^4$ square degrees) is accessible when keeping only zenith angles $<60^{\circ}$ to ensure an energy threshold $\leq$100 GeV. Assuming each pointing has a useable field-of-view of $5^{\circ}$ (about 20 square degrees) then a survey of the whole accessible sky needs about 500 different pointings. The CTA design reaches a sensitivity of 20 mCrab at 5$\sigma$ significance level in 30 mn \citep{2010arXiv1008.3703C}. Hence, a quarter of the sky could be reasonably surveyed using about a quarter of the observing time in a year (250 hr) down to a level of 20 mCrab, equivalent to the flux level of the faintest AGN currently detected at VHE energies (see \S\ref{strategy} for a discussion of the feasibility). The northern CTA site, with more emphasis on a low energy threshold, would be well suited to survey the extragalactic sky and detect faraway VHE sources, which are significantly affected by gamma-ray absorption on the extragalactic background light (EBL). However, both southern and northern sites are required for access to the whole extragalactic sky.

Such an ``all-sky" blind survey has never been done by Cherenkov arrays and would improve over current or planned VHE surveys of comparable extent in area. Key scientific questions that such a survey could impact include a census of VHE emitting Active Galactic Nuclei (AGN), looking for emission from radio galaxies cores, kpc jets, low luminous AGN or nearby galaxies. Blind surveys avoid possible biases (with the caveat that highly variable sources may be missed). For instance, there are blazars such as 1ES 0229+220 that are detected by current IACTs but not by {\it Fermi}/LAT.  The number of such sources, which bring important constrains on the intergalactic magnetic field strength \cite{ner10}, is expected to increase with CTA. Most importantly, such a survey could uncover new, unsuspected classes of extragalactic VHE sources (dark accelerators). Such a survey could constrain the density in the Galactic halo of cloudlets, cold and dense clumps of material that may constitute a sizable fraction of baryonic matter, which are mostly invisible but for their gamma-ray emission from cosmic ray interaction \cite{2004ApJ...610..868O,2005ApJ...621L..29T}. Blind search for annihilation in dark matter subhalos of the Milky Way \citep{2008Natur.454..735D} can be performed without any {\it a priori} association with an astrophysical object (dwarf galaxy, Galactic Center). Conservative estimates \citep{2011PhRvD..83a5003B} show that a 1/4th sky survey could obtain the best constraints on dark matter in the TeV regime, with a sensitivity to the natural value of the annihilation cross section for thermally-produced dark matter. Diffuse emission can be probed on scales of several degrees, constraining the distribution of cosmic rays in our Galaxy, notably the presence of a Galactic wind \cite{1991A&A...245...79B,2008ApJ...674..258E}. The survey could be correlated with all-sky maps obtained by ultra-high energy cosmic ray and high energy neutrino experiments. Localized anisotropies in the arrival directions of multi-TeV charged particles \citep{2006Sci...314..439A,2008PhRvL.101v1101A,2011arXiv1105.2326I}, could also be investigated. Limits on this program are that the sensitivity to diffuse emission declines with source size and the angular scales that can be probed cannot be much larger than the field-of-view because of background uncertainties.

\subsection{Multi-wavelength context}
\begin{figure}
\resizebox{\hsize}{!}{\includegraphics{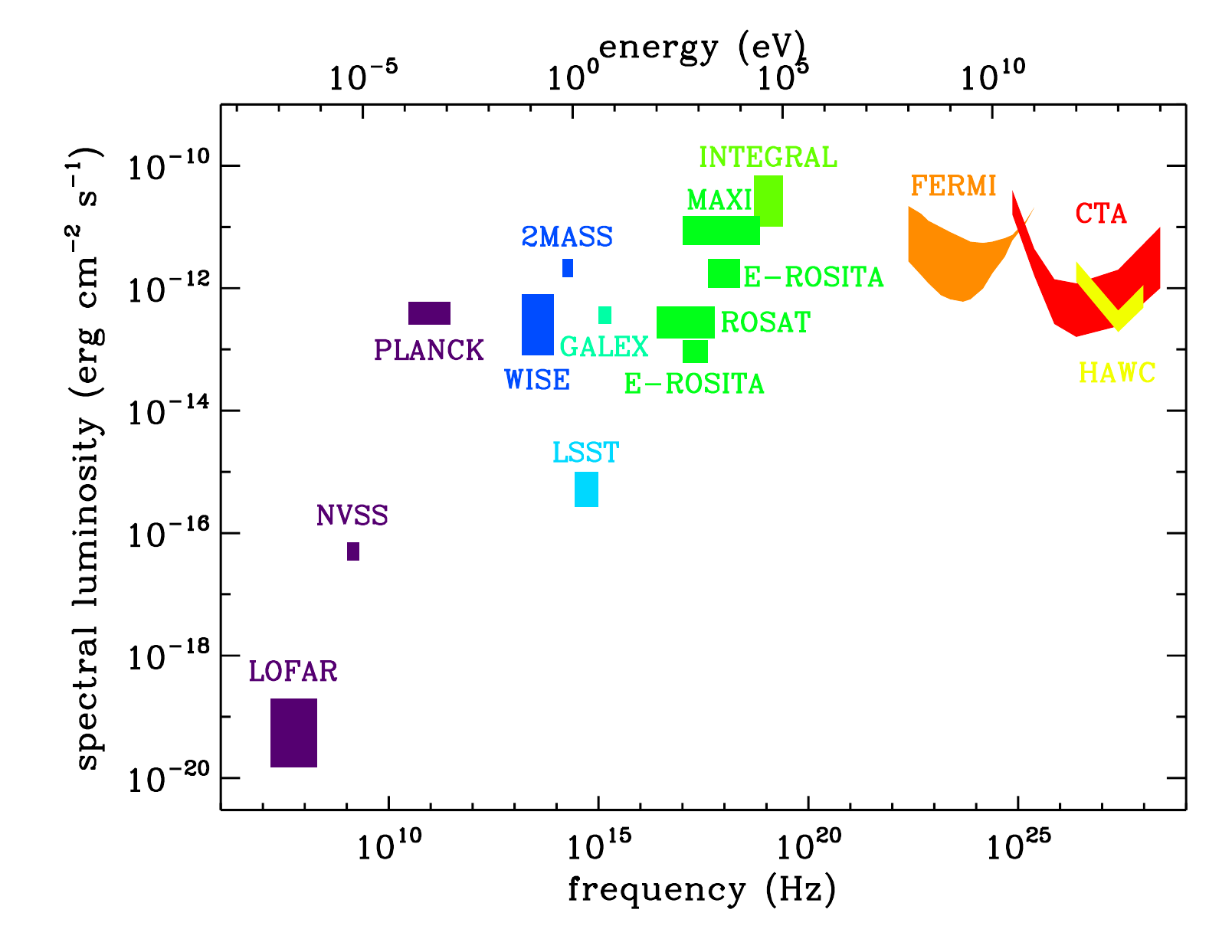}}
\caption{CTA surveys in their multi-wavelength context. The orange region corresponds to the  {\em Fermi}/LAT differential sensitivity after 10 years of observations, in the Galactic Plane (upper envelope) and outside the Galactic Plane (lower envelope). The yellow region corresponds to the HAWC sensitivity$^2$ after 1 and 5 years.  The CTA sensitivity region corresponds to pointings of one (upper envelope, also comparable to the sensitivity achieved in the HESS Galactic Plane survey) and 10 hr (lower envelope), illustrating achievable sensitivities for the all-sky and Galactic Plane surveys. Boxes show roughly the sensitivity of other surveys covering at least 25\% of the sky.\label{mwl}}
\end{figure}

There are 1873 sources in the second {\em Fermi}/LAT catalogue corresponding to an average of one HE source per 20 square degrees {\em i.e} one in every CTA field-of-view. Hence, CTA survey observations at VHE nicely complement the HE {\em Fermi}/LAT observations. Indeed, a {\em targeted survey} to {\em Fermi}/LAT sources is an alternative to a blind all-sky survey, albeit one that introduces biases (see \S\ref{fermi}). The one year {\em Fermi}/LAT point source sensitivity is $\geq 10^{-12}$ erg\,cm\,$^{-2}$\,s$^{-1}$ around 1 GeV. Diffuse emission in the Galactic Plane  worsens the sensitivity at low energies by a factor as large as $\approx 10$ near the Galactic Center. The limiting flux in the second {\em Fermi}/LAT catalogue is $5\times 10^{-12}$ erg\,cm\,$^{-2}$\,s$^{-1}$ \cite{2011arXiv1108.1435T}. Above 30 GeV, the sensitivity is  $\geq 10^{-11}$ erg\,cm\,$^{-2}$\,s$^{-1}$. Any CTA survey would go deeper than {\em Fermi}/LAT above $\approx$ 50 GeV, even taking into account that {\em Fermi}/LAT could have accumulated nearly 10 years of observations of the HE sky by the time CTA enters operation (Fig.~\ref{mwl}). The number of {\em Fermi}/LAT sources detectable by CTA based on an extrapolation of their HE spectra is discussed below (\S\ref{fermi}). CTA surveys would also complement all-sky X-ray monitoring by MAXI, which associates degree angular resolution to mCrab sensitivity (in one week) in the 0.5 - 30 keV range, and LOFAR in the low frequency radio bands. 

Figure~\ref{mwl} shows how the proposed CTA surveys complement other wide area surveys across the electromagnetic spectrum. For a given instrument, the sensitivities depend on many parameters including  wavelength, pointing direction, extension etc. They have been roughly translated as boxes in a $\nu F_\nu$ diagram: this plot does not claim to provide a highly accurate view of the respective sensitivities. For example, the LOFAR box corresponds to a survey sensitivity of 0.1 mJy from 15 MHz to 200 MHz. The HAWC integrated point source sensitivity is for one year and five years (HAWC website). The {\em Fermi}/LAT box covers the 10-year point source integrated sensitivity within the Galactic Plane (upper envelope), where diffuse emission is strong, and outside the GP (bottom envelope) \cite{2009ApJ...697.1071A}\footnote{The curves have been derived from \url{http://www.slac.stanford.edu/exp/glast/groups/canda/lat_Performance.htm}  assuming a $\sqrt{\rm time}$ scaling}. The upper envelope of the CTA region assumes a sensitivity $\approx$ 20 mCrab, achievable for the all-sky survey. Note that this limit corresponds roughly to the sensitivity of the HESS Galactic Plane survey \cite{2008ApJ...679.1299F}. The bottom solid line assumes 10 hr per pointing, achievable for the Galactic Plane survey. CTA surveys bring significant improvements and probe $\nu F_\nu$ fluxes comparable to the best X-ray or IR all-sky surveys.

Figure~\ref{mwl}  can be compared to the typical spectral energy distribution (SED) of various types of objects to investigate the best wavelengths and strategies for detection. A pulsar wind nebula 50 times fainter than the Crab would be detected in the CTA Galactic Plane survey yet would be missed by HAWC or {\em Fermi}/LAT. A SNR like RX\,J1713.6-3946 is detectable in the Galactic Plane survey essentially anywhere in the Galaxy but the faintest are missed by other gamma-ray instruments. However, a faint SNR such as SN\,1006 is barely detected in the Galactic Plane survey. A gamma-ray binary like HESS J0632+057 can be detected at close to 10 times fainter fluxes (3 times further away) in the Galactic Plane survey (with the caveat that the source is known to be variable on yearly timescales). Interestingly, such an object would not be detected in surveys at any other wavelength: CTA is the prime instrument to discover such sources. Comparing to the blazar sequence SEDs, the all-sky survey is most interesting for the extreme-peaked blazars which have very hard, faint fluxes in the {\em Fermi}/LAT range. Again, CTA is the instrument of choice for discoveries, with VHE observations drawing attention to candidate AGNs that may otherwise escape notice \citep{2011A&A...529A..49H}. A detailed study is presented in \S\ref{blazar}.

\section{Survey feasibility, performance and implementation\label{pm}}
The feasibility, performance and implementation of the proposed CTA surveys depends on several technical issues: the array configuration, field-of-view, the off-axis performance, the sensitivity to point sources and extended sources, the pointing mode, the operational mode (full array vs sub-array), etc. Particularly critical numbers are the field-of-view ($>5^{\circ}$) and sensitivity ($<10$mCrab for 5$\sigma$ in one hour). We report here on the detailed studies that have been carried out to quantify the sensitivity using the latest CTA responses. The responses are calculated for various array configurations differing in the number, size and position of the telescopes and labelled by letters A, B, C, etc. The Monte Carlo simulations and array configurations that lead to the CTA responses used here are described elsewhere in this volume.

\subsection{Simulation tool\label{tool}}
The evaluation of different survey strategies and their achieved source sensitivity was done in a realistic way, by taking into account the anticipated CTA performance as well as the requirements of the data analysis stage. For that purpose, we used the ctools, a set of tools built from GammaLib, an open-source C++ library that contains all the functionalities needed for the high-level analysis of astronomical gamma-ray data\footnote{see \url{http://cta.irap.omp.eu/ctools/}}. 	
In our simulations, CTA is defined by its energy-dependent effective area, point-spread function and instrumental background. These were taken from the Monte-Carlo studies of the configuration E  for most cases, but we also assessed alternative array configurations and subarrays of a few telescopes.  We used the instrument response functions (IRFs) optimized for 50h observation time in most cases, except for the all-sky surveys where short exposures of 30mn / 1 hr are involved; in the latter case, the sensitivities were computed with the IRFs optimized for 30 mn. We also tested 5 hr IRFs for the Galactic Plane survey, finding differences of less than 1 mCrab in computed sensitivity. We kept 50 hr IRFs for ease of comparison with other studies. The dependence of the background and effective area on the off-axis angle was assumed to be Gaussian in off-axis angle squared, with a standard deviation $\sigma_\mathrm{FoV}=3^\circ$ (where FoV stands for Field-of-View). This is a good approximation of the Monte Carlo simulation output for configuration E and energies around 1 TeV, where the array is most sensitive. However, this is an over/underestimate at 0.1/10 TeV (respectively) because of the dependence of acceptance on  energy and the different fields-of-view of large/small-sized telescopes. In configuration E, large-sized telescopes (most important at low energies) have $\theta_{\rm FoV}=4.6^\circ$, medium-sized telescopes have $\theta_{\rm FoV}=8^\circ$ and small-sized telescopes (most important at high energies) have $\theta_{\rm FoV}=10^\circ$. Configuration I, which also has a balanced sensitivity across the whole energy range, uses medium-sized telescopes with the same field-of-view but more small-sized telescopes with $\theta_{\rm FoV}=9^\circ$. Configuration E should be representative although other configurations may have slightly different values of $\sigma_\mathrm{FoV}$, a key parameter of this study. Regarding the point-spread function, any off-axis angle dependence has been ignored in this study, which may result in an overestimation of the angular resolution of up to a factor of 2 at the edge of the field-of-view.
	
For a specified pointing strategy (layout and observing time) and source model (position, shape, and flux), mock datasets were generated that consist of background and source events defined by their reconstructed energy and direction. No astronomical visibility constraints have been applied. Hence, the effects of the pointing zenith angle are not taken into account and only IRFs determined for a zenith angle of 20$^\circ$ were used. This implies slightly optimistic values for the effective areas or PSF compared to the average values that would be obtained in a real implementation, where observations would be taken with larger zenith angles.

The source contribution only comes from discrete sources and no Galactic diffuse emission was input at this stage. These observations were then analyzed by model-fitting using a maximum likelihood procedure for unbinned data. The Monte-Carlo spectrum for the background is fitted to the data simultaneously to a convolved model for the source, and this provides measurements for the source flux, index and extent. The sensitivity over the surveyed area for a given pointing strategy was determined iteratively by adding a test source in the field and finding for that position the source flux that leads to a 5$\sigma$ detection, defined by the obtention of a Test Statistic (TS) $>$ 25. In the process, only the source flux was fitted. All other properties like spectral index, position, and shape were fixed at their true values; this means that our sensitivity estimates are slightly optimistic. The results presented thereafter correspond to the 100 GeV-100 TeV energy range, and are expressed in terms of flux density at 0.3 TeV relative to the Crab. For a test source spectral index $\Gamma$=2.5, which is what we used in most cases, this is an actual scaling in integrated flux (assuming the Crab spectrum is a single power-law spectrum with $\Gamma$=2.5).

\subsection{Evaluation of survey pointing strategies\label{strategy}}
\begin{figure}
\centering
\resizebox{5cm}{!}{\includegraphics{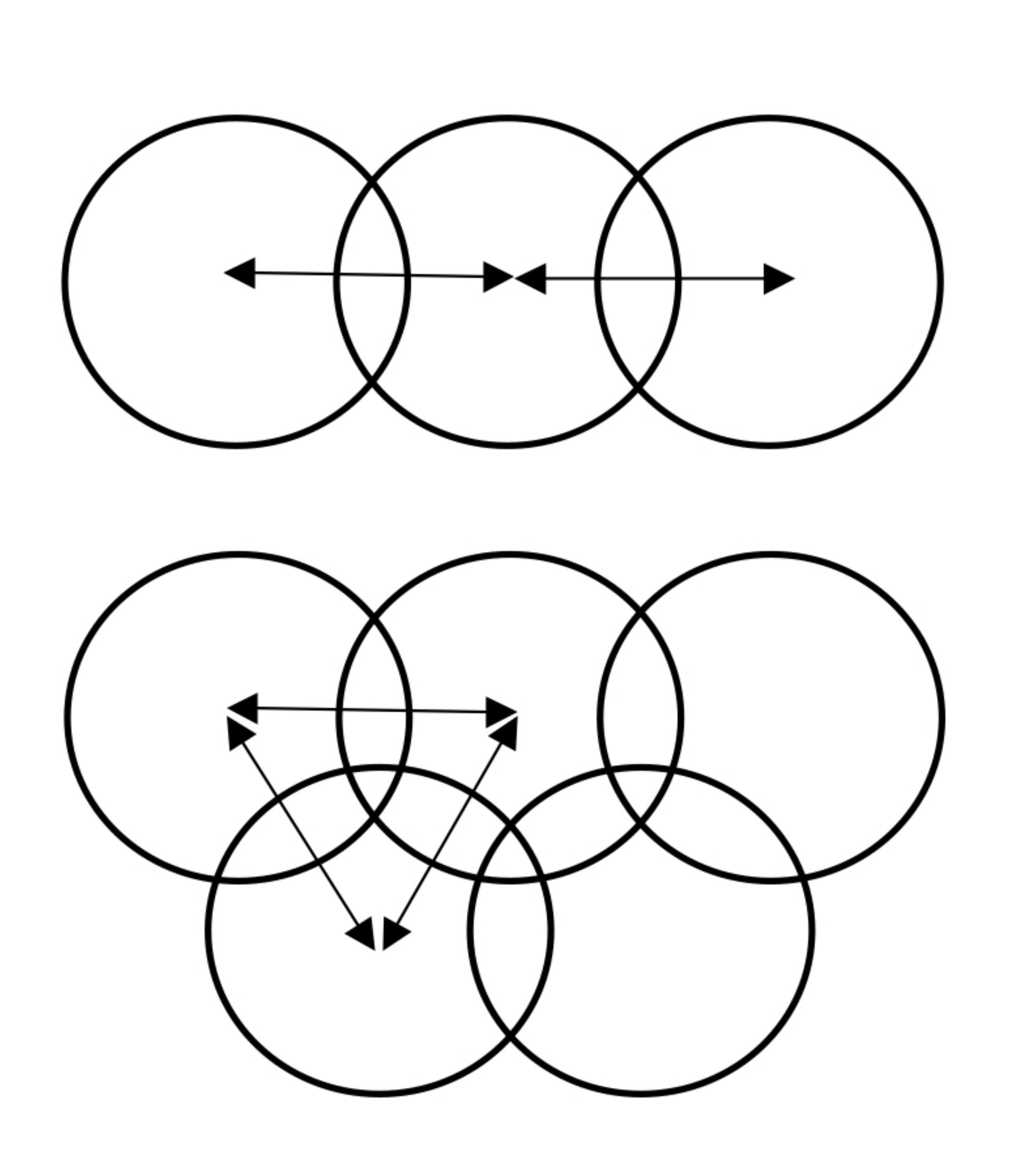}}
\caption{Tiling strategy for surveys: single row for Galactic Plane (top) or equilateral triangles for ``all-sky" (bottom).}
\label{tiling}
\end{figure}

\begin{figure}
\centering
\resizebox{\hsize}{!}{
\includegraphics{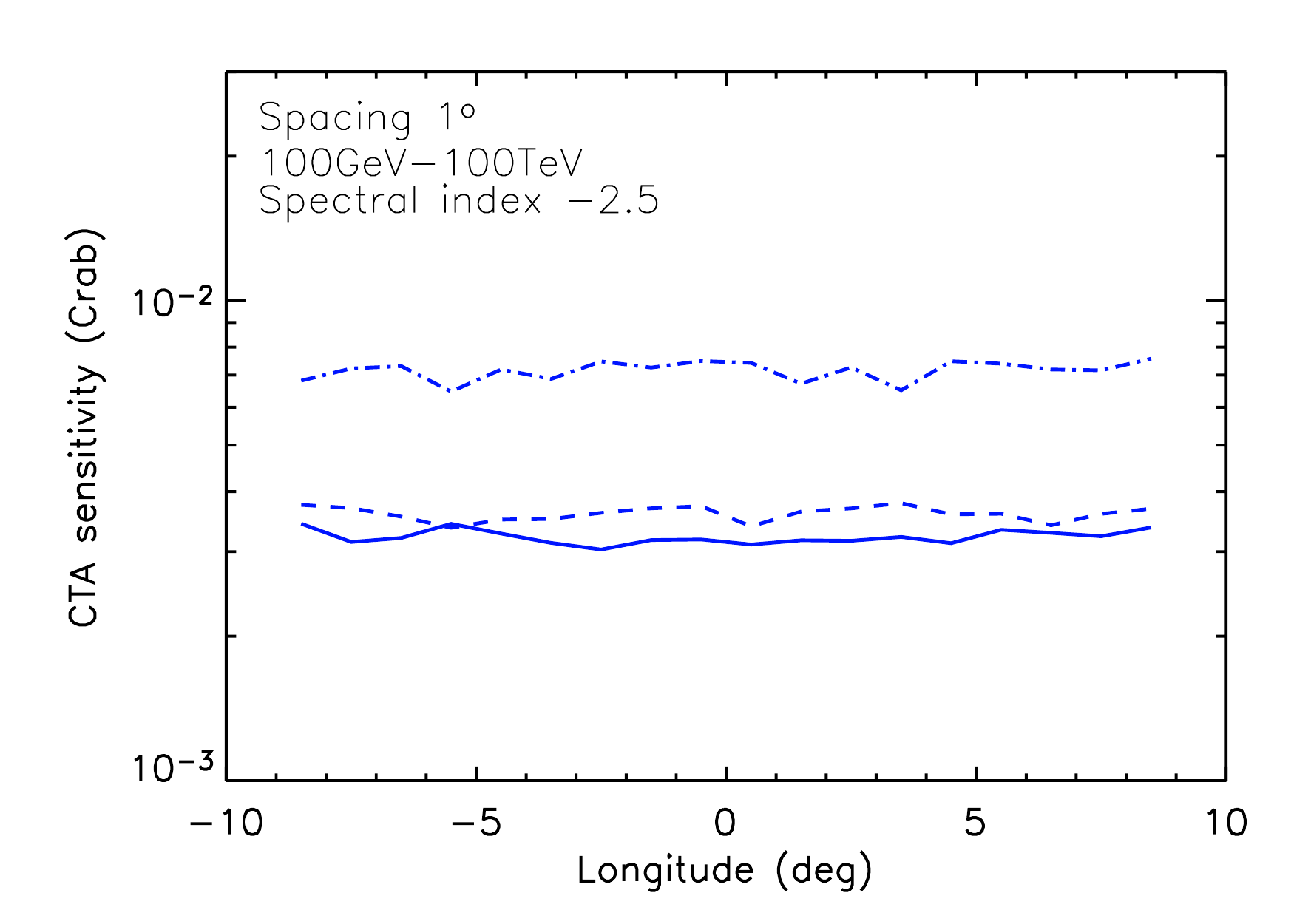}
\includegraphics{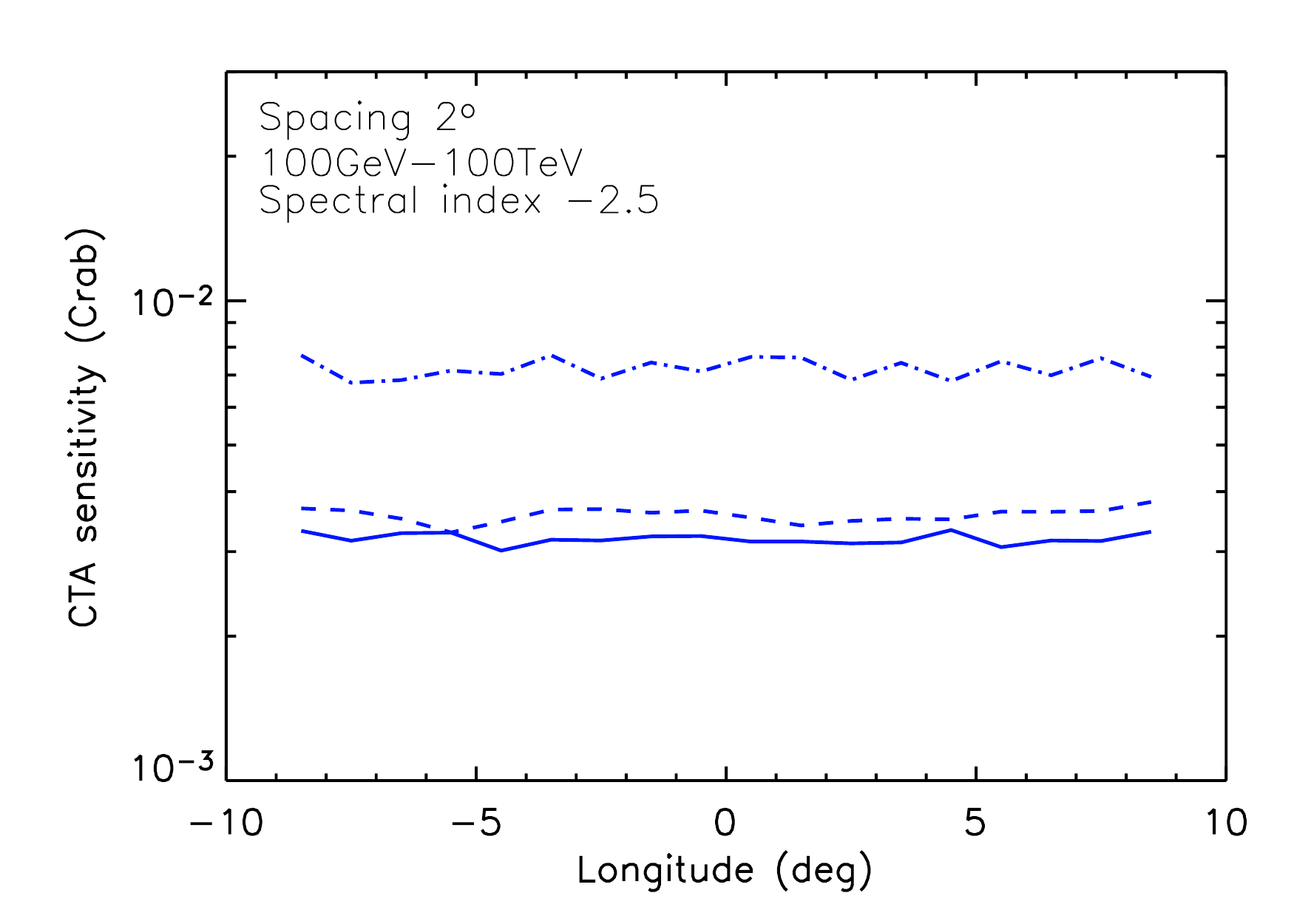}
}
\resizebox{\hsize}{!}{
\includegraphics{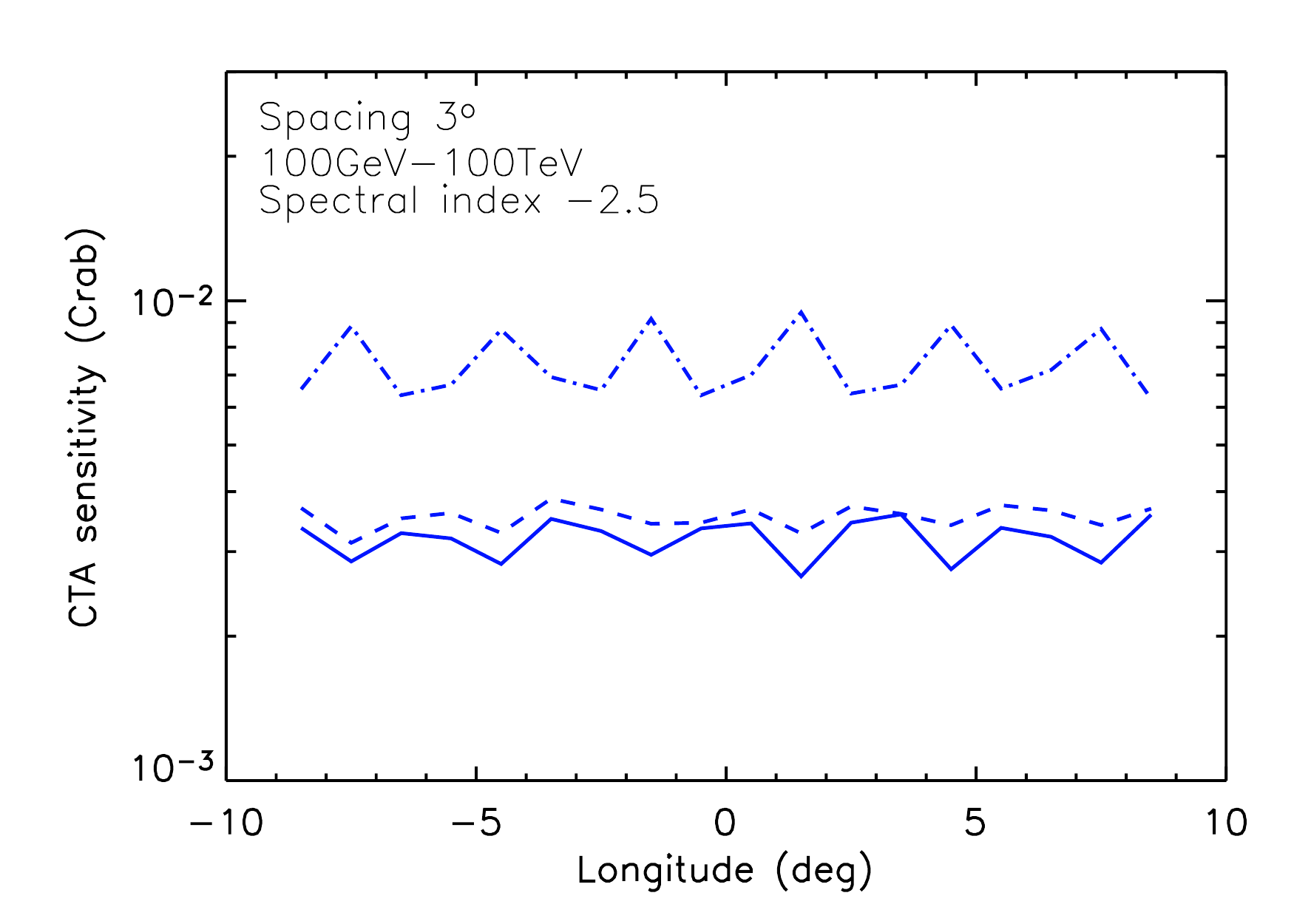}
\includegraphics{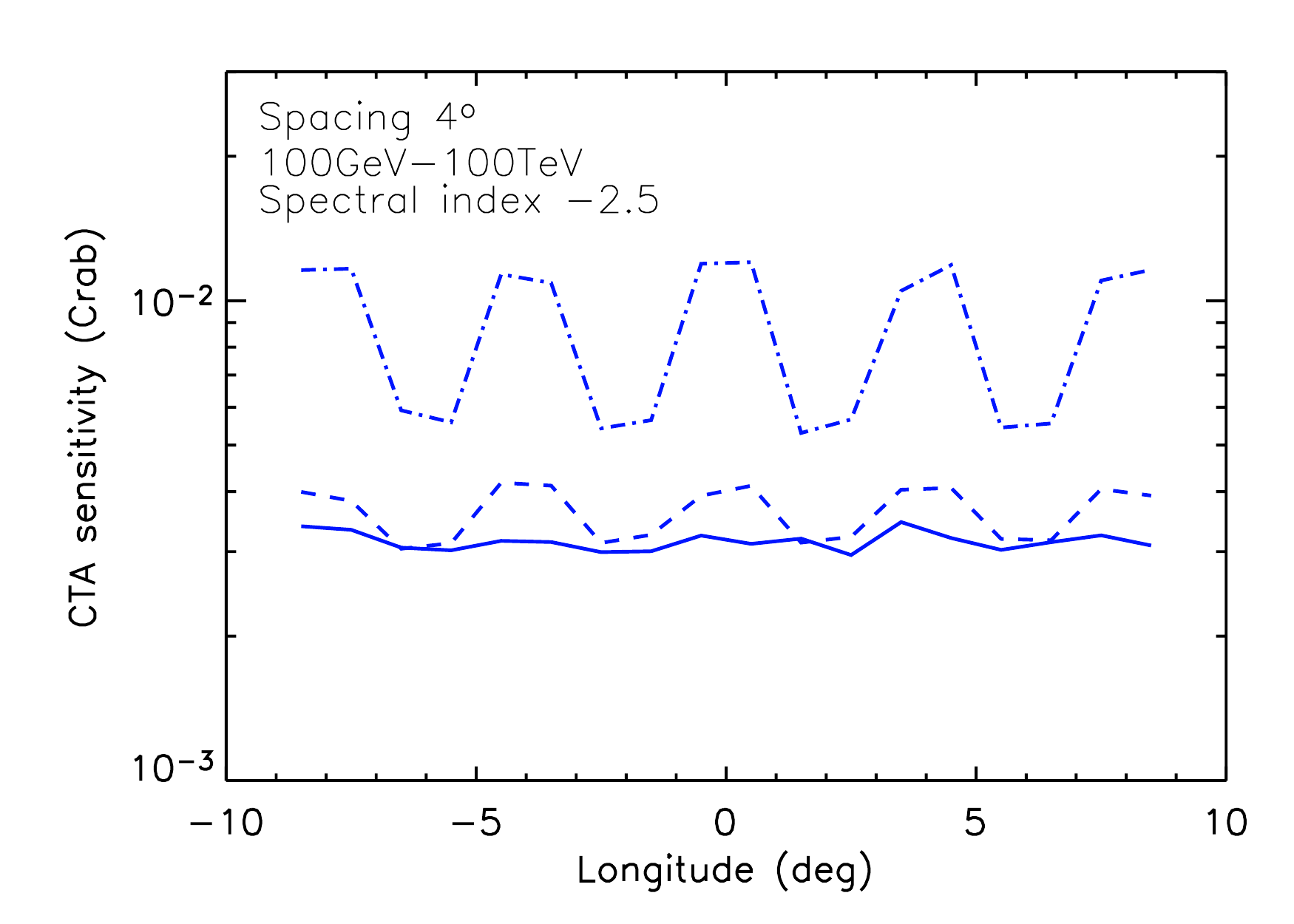}
}
\caption{Sensitivity along longitude $\ell$ to a point source with spectral index $\Gamma$=2.5 over the 100 GeV-100 TeV range, for simulated 240 hr Galactic plane single-row surveys with different pointing steps. Solid, dashed, and dot-dashed lines correspond to the sensitivity at latitudes $b$ of 0, 1, and 2$^\circ$ respectively.}
\label{steps}
\end{figure}
We examined two survey strategies to cover a given area of the sky: a single row of pointings ({\em e.g.} Galactic Plane survey) and multiple evenly-spaced rows of pointings ({\em e.g.} wide area ``all-sky" survey). For the latter option, the tiling motif is an equilateral triangle (Fig.~\ref{tiling}). Hence, the grid is uniquely characterized in both cases by the angular step between adjacent pointings. 

For the Galactic Plane survey, the objective is uniform sensitivity over the longitude range $-60^\circ \leq \ell\leq 60^\circ$ using a total observing time of 240 hr ($\approx$ 1/4th observing time in one year of operations at one CTA site, and equivalent to the initial HESS Galactic Plane survey). This area of the sky is fully accessible from the proposed southern sites for CTA (Fig.~1). We searched for the optimal longitude step in between pointings, for a single row of pointings aligned along the Galactic plane at $b=0^\circ$. Figure \ref{steps} shows the 5$\sigma$ sensitivity obtained in and slightly off the plane for various steps (the longitude range of the plot is restricted for simplicity). A sensitivity of 3 mCrab is reached within $|b|\leq 1^\circ$ for all steps $\leq 4^\circ$. Steps larger than that lead to inhomogeneous coverage along the plane, with systematic sensitivity fluctuations of about an order of magnitude for a step of $6^\circ$. The superposition of circular FoVs also leads to increasingly inhomogeneous coverage off the plane as the step gets larger. For steps up to 2$^\circ$, the sensitivity decreases with latitude, reaching 7 mCrab at $b=2^\circ$, but remains quite uniform at any given latitude. For larger steps, the sensitivity exhibits significant variations with longitude off the plane, preventing a uniform coverage beyond $1^\circ$ in latitude.

The number of pointing directions increases with decreasing step size. The sensitivity along the plane at $b=0^\circ$ is the same in all cases but the exposure time required for each pointing direction changes with step size. For spacings of 1$^\circ$, 2$^\circ$, 3$^\circ$, or 4$^\circ$, this implies 120, 60, 40, or 30 pointings of 2 hr, 4 hr, 6 hr, or 8 hr each, respectively. A value of 2$^\circ$ appears as a good compromise between latitude coverage and number of pointings. A two-row strategy with a spacing of 3$^\circ$ is an alternative for the Galactic Plane survey. It provides a more uniform coverage in latitude, at the expense of a reduced sensitivity in the midplane. As an example, a double-row survey with a spacing of 3$^\circ$ and a total time of 240 hr gives a sensitivity ranging from 4 mCrab in the plane to 5 mCrab at $b=2^\circ$, to be compared with a sensitivity ranging from 3 mCrab in the plane to 7 mCrab at $b=2^\circ$ for a single-row strategy with a spacing of $2^\circ$. The double-row strategy requires more pointings (80 instead of 60 in this specific case). The number of currently known sources drops rapidly with latitude, most being within 1$^\circ$, which tends to favor the single-row strategy with high mid-plane sensitivity. On the other hand, an extended latitudinal coverage may help for background subtraction, as it provides more off observations.

The same analysis was performed to find the optimal 2D grid for a wide-area survey, using IRFs appropriate for 30 mn exposures. Steps $\leq 2^\circ$ were found to provide nearly uniform coverage, while systematic sensitivity variations appear for greater step sizes because of the non-overlap of the circular FoVs. Quantitatively, for pointings of 1 hr each, steps of 1$^\circ$ lead to average sensitivities over the field of about 3 mCrab with deviations of $\pm$4\%. The sensitivities for steps of 2$^{\circ}$, 3$^{\circ}$, 4$^{\circ}$, or 5$^{\circ}$ are 6 mCrab, 10 mCrab, 14 mCrab, or 24 mCrab with deviations of 3\%, 12\%, 9\%, or 47\%, respectively. Using pointings of 0.5 hr each, the typical duration of an IACT run, gives average sensitivities for steps from 1$^{\circ}$ to 5$^{\circ}$ of  5 mCrab, 9 mCrab, 15 mCrab, 22 mCrab or 36 mCrab with the same deviations as before. These results show that pointing steps $\geq 5^\circ$ would not provide a homogeneous sampling of the sky. If sensitivity variations at the 10\% level can be considered as acceptable, a uniform survey of about 1/4th of the sky can be done at the 22 mCrab level ($\geq 100$ GeV) with a 4$^\circ$ evenly-spaced grid of about 740 pointings of 0.5 hr each (370 hr compared to the 1000 hr available in a year of operations).  We find that the sensitivity above 1 TeV is somewhat better (38 mCrab) than the HAWC one-year sensitivity above the same energy threshold (50 mCrab). A CTA all-sky survey offers better angular resolution and a much lower energy threshold, a major advantage for extragalactic sources, which tend to be soft. On the other hand, the CTA all-sky survey cannot be expected to offer much variability information because each position on the sky is visited a couple of times at most (see Fig.~\ref{steps}). HAWC remains the instrument of choice to explore the variable and transient sky. 

\subsection{Effect of spectral index and source size\label{index}}
The sensitivities given so far hold for point sources with a spectral index $\Gamma$=2.5 integrated over the 100 GeV-100 TeV energy range. Since TeV sources such as SNRs and PWNe are expected to have a distribution in spectral index and size, we investigated the effect of these parameters on the anticipated sensitivities. We performed that study for the two survey strategies identified above as promising scenarios: a single-row Galactic Plane survey with a 2$^\circ$ spacing and 60 pointings of 4 hr each (giving an effective exposure time $\approx$ 8 hr at each location on the sky), and a multiple-row all-sky survey with a 4$^\circ$ spacing and 740 pointings of 0.5 hr each (giving an effective exposure time $\approx$ 0.5 hr per location). Only the flux of the test source is fitted, while all other source parameters are fixed to their true values.

For the Galactic Plane survey, the sensitivity to a point source with $\Gamma$=2.0, 2.5, or 3.0 is about 1 mCrab, 3 mCrab or 5 mCrab respectively. For the all-sky survey, the sensitivity as function of $\Gamma$ becomes 10 mCrab, 22 mCrab and 25 mCrab (respectively). The higher sensitivity to hard sources is due to the combined effects of lower instrumental background, sharper point-spread function, and larger effective area at higher energies. For the Galactic Plane survey, the sensitivity to a source with $\Gamma$=2.5 and a disk-like uniform intensity distribution of radius 0.0$^{\circ}$, 0.1$^{\circ}$, 0.2$^{\circ}$, or 0.3$^\circ$ is 3 mCrab, 5 mCrab, 9 mCrab, and 12 mCrab (respectively). The numbers for the all-sky survey are 22 mCrab, 28 mCrab, 42 mCrab, and 58 mCrab (respectively). This comes from the signal being increasingly spread in the instrumental background.
\begin{figure}
\centering
\resizebox{\hsize}{!}{\includegraphics{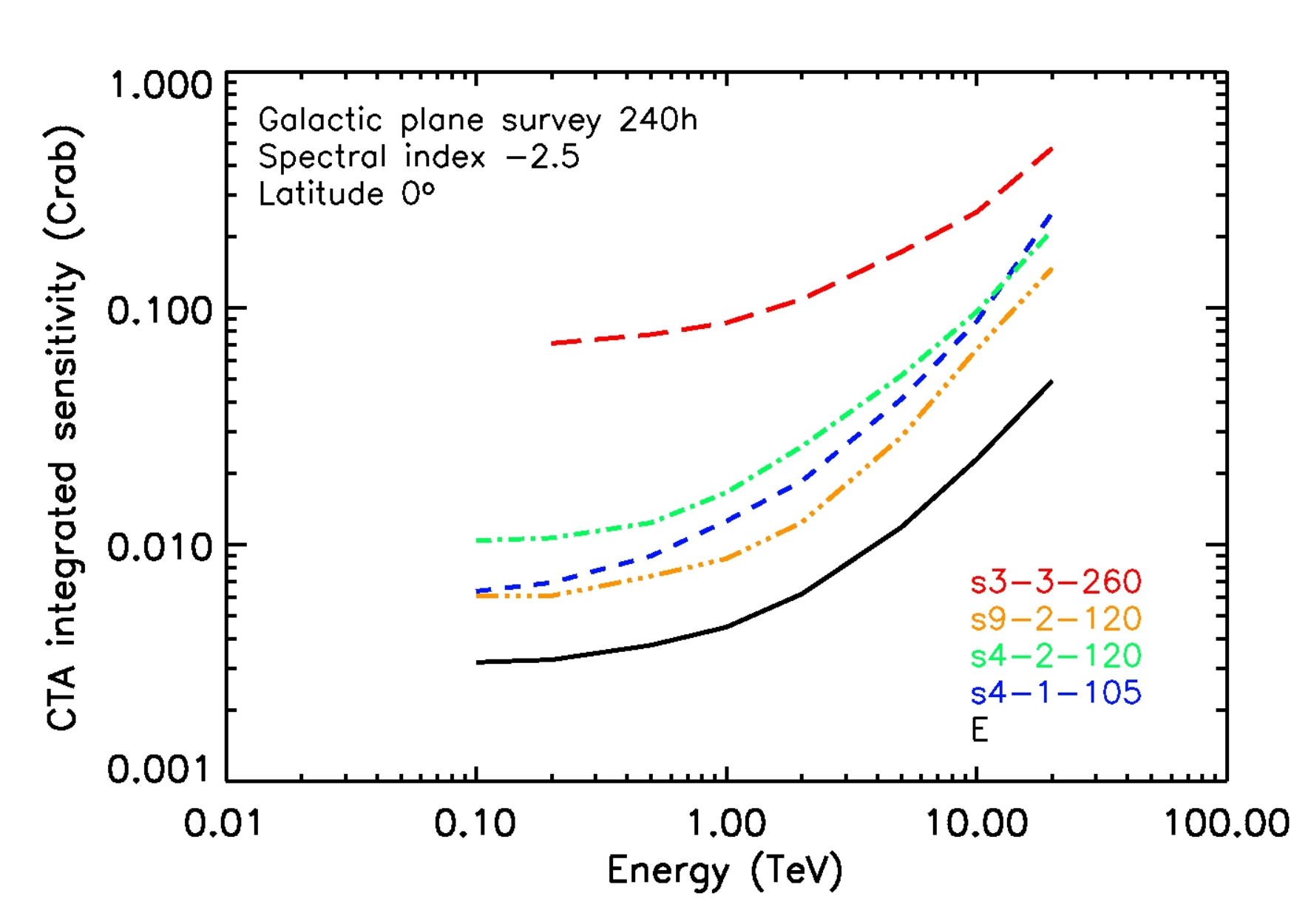}}
\caption{CTA integrated sensitivity to a point source at $b$=0$^\circ$ with spectral index $\Gamma=2.5$ in a 240 hr survey of the Galactic Plane using either the full array in configuration E  or subarrays. The integrated sensitivity is given as a function of the threshold energy.}
\label{subarray} 
\end{figure}

\subsection{Surveys with subarrays\label{sub}}
The large number of individual telescopes involved in CTA can provide important flexibility in operation and the ability to pursue several scientific objectives in parallel by using subsets of the entire array. We assessed the survey performances of several subarray configurations\footnote{In the subarray designations, the first number refers to the number of telescopes in the subarray, the second to the type of telescope --- 1 for LST, 2 for MST, 3 for SST --- the last number is the separation in meters between telescopes, e.g. s4-2-120 is a HESS-like configuration of 4 medium-sized telescopes with a separation of 120 meters.} studied by the Monte-Carlo group: s4-1-105, s4-2-120, s9-2-120, and s3-3-260. The first and last ones can be exactly implemented in array configuration E, but the other two cannot and were considered only to see the impact of using subarrays of medium-sized telescopes only.

The energy-dependent effective area, on-axis point-spread function and instrumental background for each subarray were obtained from the Monte-Carlo studies. The off-axis dependences are the same as those assumed above for the entire array E. We emphasize again that this approximation needs to be improved at low/high energies, especially when using only large/small telescopes. In the context of surveys with subsets of telescopes of the same kind, this means that the pointing step and duration should be adjusted for each subarray (for instance, large-sized telescopes with a small field-of-view may require a smaller step to get a uniform sensitivity coverage).
	
We simulated a survey of the Galactic Plane with the strategy outlined in the previous subsection using different subarrays. Fig.~\ref{subarray} shows the integrated sensitivities above a given energy achieved for each array/subarray along the Galactic Plane. These hold for a point source with spectral index $\Gamma$=2.5. The integrated sensitivity above 100 GeV is lower by a factor of 2 ($\approx $ 6 mCrab) when using only the 4 large-sized telescopes (s4-1-105) instead of the full array E. The sensitivity loss worsens as the threshold energy increases since the subarray lacks the small-sized telescopes that provide high-energy sensitivity. Note that surveys with this subarray are likely to require a much greater number of pointings than assumed here to cover the same area because of the narrower FoV of large-sized telescopes (4.6$^\circ$ compared to 8$^\circ$ for the medium-sized telescopes). Using only 3 small-sized telescopes (s3-3-260) leads to a sensitivity drop by at least 20 compared to the performance of the full array, at a level of 70 mCrab at best. A HESS-like array of 4 medium-sized telescopes (s4-2-120) provides a sensitivity of about 10 mCrab above 100 GeV, and this decreases to 6 mCrab for 9 medium-sized telescopes (s9-2-120). In both cases, the sensitivity difference compared to array E remains approximately constant over the energy range considered here. However, the angular resolution of a subarray is not as good as that of the full array (68\% containment radius of $\approx 0.08^\circ$ for subarray s4-2-120 compared to about 0.05$^\circ$ at 1 TeV for the full array E), which may worsen the issue of source confusion in the Galactic Plane (\S\ref{mr}). 

\subsection{Surveys in divergent mode\label{divergent}}
An alternative strategy for a survey with CTA is to move from convergent to divergent pointing of the telescopes. Considering a $\sim$ 25 medium-sized telescope subarray, the angles between telescope pointing directions can be adjusted such that a 20$^\circ \times 20^\circ$ patch of sky can be covered with an average of 2-3 telescopes observing a given event. This situation can be approximated by considering the sensitivity of the HESS-like subarray s4-2-120 (the sensitivity is essentially set by the telescope multiplicity) but with uniform exposure over a 20$^\circ \times 20^\circ$ FoV. The small-sized telescope array could be used to cover the same FoV with increased telescope multiplicity, or covering a wider FoV due to the increased number and FoV of the telescopes. For the large-sized telescopes only a modest increase in sky coverage is possible due to the small number of telescopes. Toy model simulations suggest that the overall survey depth achieved by CTA is rather flat as a function of the degree of divergence. This mode sacrifices precision, the energy and angular resolution are comparable to HESS, for instantaneous sky-coverage.

The large area surveyed with each pointing of the array greatly helps an all-sky survey. Assuming the sensitivity of the s4-2-120 subarray, an exposure duration of $\approx$ 4 hr is required to reach about 20 mCrab over 100 GeV-100 TeV for a source with spectral index 2.5. To cover 1/4th of the sky, about 25 pointings of the 20$^\circ \times 20^\circ$ enlarged FoV are required, which makes a total of about 100 hr. This is nearly 4 times less than the total time for convergent pointing with the full array E at the same sensitivity of 20 mCrab (\S\ref{strategy}). The time to complete the survey is smaller in divergent mode as long as each pointing covers more than 10$^\circ \times 10^\circ$ uniformly with  the s4-2-120 subarray sensitivity, a goal that appears quite achievable using a dozen mid-sized telescope with a FoV of 8$^{\circ}$. A large extension in latitude $b$ is not required for the Galactic Plane survey (see \S2.1). In this case, pointings diverging only in longitude could be considered instead of divergent pointings covering a 20$^\circ \times 20^\circ$ patch. However, the gain in observing time compared to successive convergent pointings may not be as great for such a unidimensional survey.

Divergent mode also appears promising in the search for transient phenomena. The successive visits required to build up sensitivity in a targeted patch of extragalactic sky provide chances to detect sources flaring at $\geq 60$ mCrab, based on the sensitivity for detection of a point source in a single visit. The visits can be spread out to probe various timescales. For example, four visits can be divided into two visits per night on consecutive nights to probe hour to day timescales. Two additional visits can be scheduled the following week and another two the following month, allowing for detection of variability on longer timescales while ensuring the total number of visits (8) is sufficient to reach the survey sensitivity goal for steady sources. Such a program is observationally feasible in principle, although we have not studied in detail its practical implementation. 

Divergent pointing offers clear advantages in terms of variability studies and investment in observing time. However, divergent modes require non-standard analysis with possible complications to {\em e.g.} background estimation since each telescope observes a slightly different direction on the sky. Further studies are being carried out to assess precisely the potential of this observing mode.

\section{Source population accessible in surveys}
The rationale for surveys depends largely on the ability of the observations to cover known populations of sources. Several studies have been carried out to quantify the numbers of detections expected from surveys for different populations, extrapolating from current knowledge: the population of blazars in the extragalactic sky (\S\ref{blazar}), the overall population of {\em Fermi}/LAT sources (\S\ref{fermi}), the population of SNRs and PWNe in the Galactic Plane (\S\ref{mr}).  A full simulation of the CTA Galactic Plane Survey was carried out and is discussed in \S\ref{sim}. The blazars and {\em Fermi}/LAT sources have been considered as point sources for CTA. Spatially-extended VHE emission has been taken into account for the Galactic sources.
	
\subsection{Blazars in a wide area survey\label{blazar}}
Blazars are the dominant population in the extragalactic gamma-ray sky: almost all of the extragalactic sources detected by EGRET and {\it Fermi}/LAT are blazars \citep{har99,abd10_catalog}. Current IACTs have already found more than 40 blazars out of 100 VHE sources, up to a redshift $z=0.536$. The number of VHE blazars and maximum detected redshift will increase with CTA. A population study of VHE blazars with CTA would provide keys to understanding AGN populations, high-energy phenomena around supermassive black holes, the cosmological evolution of AGN and the extragalactic background light (EBL) \citep{cta_agn}.

The prospects for future blazar surveys by CTA are considered here. To evaluate the potential of surveys for such studies, we use a model of the blazar gamma-ray luminosity function and spectral energy distribution (SED) to predict the expected number and distributions of physical quantities of VHE blazars in a future CTA sky survey. We use the new blazar luminosity function presented  in \citep{ino09,ino10}, which takes into account the blazar SED sequence \citep{fos98,kub98} and is in  agreement with the EGRET and {\it Fermi}/LAT data \citep{ino10,ino11a}. Standard cosmological parameters are adopted, $(h, \Omega_M , \Omega_\Lambda) = (0.7, 0.3, 0.7)$.

\begin{figure*}
\centering
\includegraphics[width=140mm]{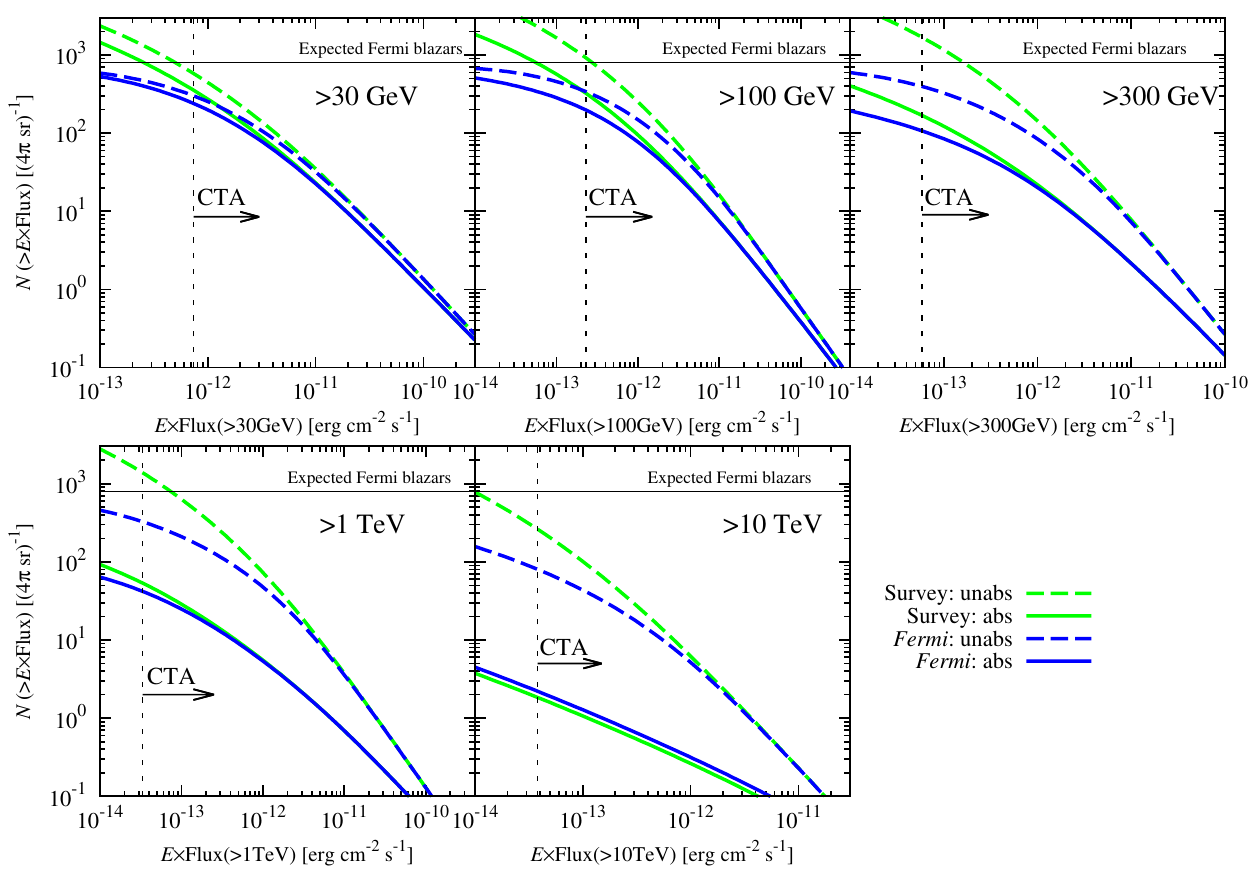}
\caption{Expected cumulative source counts as a function of the integral gamma-ray flux of VHE blazars. The five panels correspond to different photon energies, as indicated in the panels. Green curves correspond to a blank field all-sky survey, blue curves for a follow-up of {\it Fermi} blazars (assuming a {\em Fermi}/LAT sensitivity limit of $3\times10^{-9} $ photons cm$^{-2}$ s$^{-1}$ above 100 MeV). The horizontal thin solid line is the total expected number of blazars above the {\em Fermi}/LAT sensitivity. Solid curves include EBL absorption, dashed curves do not. The CTA 5$\sigma$, 50 hr detection limit with array E is also shown. The blue solid curve in the panel of 10 TeV is shifted upward artificially by a factor of 1.2 for the purpose of presentation, because the blue solid and green solid curves totally overlap with each other.}
\label{fig:count}
\end{figure*}

Following  \cite{ino10}, we consider the case for a blind sky survey (all-sky survey) and a targeted survey ({\em e.g.} to {\em Fermi}/LAT-selected targets, see also \S\ref{fermi} below).  The dependence of source counts on the key parameters of the survey --- Field-of-View (FoV), observing time per FoV, $t_{\rm FoV}$,  total observing time, $t_{\rm obs}$ --- is analytically estimated as follows. The total source counts $N(>F)$ above a certain flux limit $F$ is given as
\begin{equation}
N(>F) = N_0\left(\frac{F}{F_0}\right)^{n} \times A_{\rm obs}(t_{\rm obs}),
\label{eq:Nf}
\end{equation}
where $N_0$ is the expected cumulative source counts per one square degree down to a flux limit $F_0$, $n$ is the slope index of the cumulative source distribution, and $A_{\rm obs}(t_{\rm obs})$ is the total survey area given by 
\begin{equation}
A_{\rm obs}(t_{\rm obs}) = A_{\rm FoV} \frac{t_{\rm obs}}{t_{\rm FoV}},
\end{equation}
where $A_{\rm FoV} = \pi \theta_{\rm FoV}^{\, 2}/4$ and $\theta_{\rm FoV}$ is the FoV in degrees.

When the flux limit depends on the inverse square root of $t_{\rm FoV}$, Eq. \ref{eq:Nf} can be rewritten using a reference flux limit $F_0$ for observing time $t_0$ as
\begin{eqnarray}
N[>F(t_{\rm FoV})] &=& N_0\left(\frac{t_{\rm FoV}}{t_0}\right)^{-n/2}\times \pi \frac{\theta_{\rm FoV}^{\, 2}}{4}\frac{t_{\rm obs}}{t_{\rm FoV}}\\
&\propto& t_{\rm obs}\, \theta_{\rm FoV}^{\, 2}\,  t_{\rm FoV}^{-(1+n/2)}.
\end{eqnarray}
When sources are uniformly distributed in the Euclidean universe, $n= -1.5$. Then, 
\begin{equation}
N[>F(t_{\rm FoV})]\propto t_{\rm obs}\, \theta_{\rm FoV}^{\, 2} \, t_{\rm FoV}^{\, -0.25},
\end{equation}
clearly showing the importance of as large a FoV as possible and  favoring a shallow survey to cover as wide an area as possible within a limited observing time.

Figure \ref{fig:count} shows the cumulative source count distributions in the entire sky above five energy thresholds (30 GeV, 100 GeV, 300 GeV, 1 TeV, and 10 TeV) from more detailed calculations \citep{ino10}. The model predicts $\approx$ 800 blazar detections with {\em Fermi}/LAT, which agrees well with observed numbers (see \citep{2011arXiv1108.1420T} and \S\ref{fermi}).  The expected detections for a follow-up of these {\em Fermi}/LAT blazars (a targeted survey instead of a blind survey) is also shown. Absorption of blazar spectra by the EBL was taken into account using \citep{EBL_Franceschini}. For high limiting fluxes there is little difference in count numbers between all-sky and targeted surveys. An all-sky survey is favored over a targeted survey in terms of number of detections as the limiting flux gets closer to the 5$\sigma$/50 hr limit, but such an exposure is unrealistic for a wide survey limited in time. However, even if the number of detections is similar, note that blind surveys and targeted surveys are not sensitive to the same classes of sources ({\em e.g.} the case of 1ES 0229+220 discussed above).

Table \ref{tab:count_7} show the expected source counts with 250 hr of total observing time for 0.5 hr (total survey area 19000 square degrees), 1.0 hr (9600 sq. degrees), 5 hrs (1900 sq. degrees), and 50 hrs (190 sq. degrees) per FoV in the case of 7$^{\circ}$ FoV and array configuration I. Sensitivities for various observational time per FoV are calculated by using internal CTA tools. Serendipitous discoveries of blazars are favored by wider, shallower surveys. CTA can be expected  to detect $\approx 20$ blazars with a 250 hr blank survey. 

\begin{table}[tbh]
\begin{center}
\begin{tabular}{ccccc}\hline
area (deg$^{2}$) &19000  & 9600  & 1900 & 190 \\
exposure (hr/FoV) &0.5 & 1 & 5 &50 \\
\hline
$>$30 GeV& 26 & 19 & 7.5 & 1.7\\
$>$100 GeV& 25 & 18 & 7.2 & 1.7\\ 
$>$300 GeV& 14 & 9.1& 4.0& 0.87\\ 
$>$1 TeV& 4.3& 2.9 & 1.2& 0.28\\ 
\hline
\end{tabular}
\end{center}
\caption{Expected blazar source counts for CTA all-sky survey (total time 250 hr) with sensitivity of array I assuming a FoV of 7$^{\circ}$ and various energy thresholds.}
\label{tab:count_7}
\end{table}

\subsection{{Fermi}/{\rm LAT} sources with CTA\label{fermi}}
The Second {\em Fermi}/LAT catalog (2FGL) represents the most complete list of sources in the GeV sky to date. To assess the potential of CTA surveys, the reported 2FGL spectral parameters for the 1873 sources were extrapolated to the very high energy range (15 GeV -- 300 TeV).  We used the integral flux from 1 to 100 GeV in ph cm$^{-2}$ s$^{-1}$ units (F1000) and spectral index $(\Gamma)$ furnished by the 2FGL catalog. For each individual source, we adopt the corresponding power-law or LogParabola parameters prescribed in \cite{2011arXiv1108.1435T}. Once the extrapolated flux is fixed, it is weighted with the simulated CTA effective area for different telescope configurations. Actual statistical significances were calculated using Eq. 17 in \cite{LiMa},  assuming $N_{\rm on}$ (on region) to be the number of source photons plus the number of photons from the background, $N_{\rm off}$ fixed at the background rate (off-region) and the number $\alpha$ given by the ratio of the sizes of the two regions, the ratio of the exposure times and the respective acceptances. For simplicity, 5 off-regions for each on-region observation and a 5\% systematic error were considered \citep{2010arXiv1008.3703C}. A detection must exceed a significance above $5\sigma$ \emph{and} a signal over 5\% of the background.

\begin{figure}
\hfil
\includegraphics[width=\linewidth]{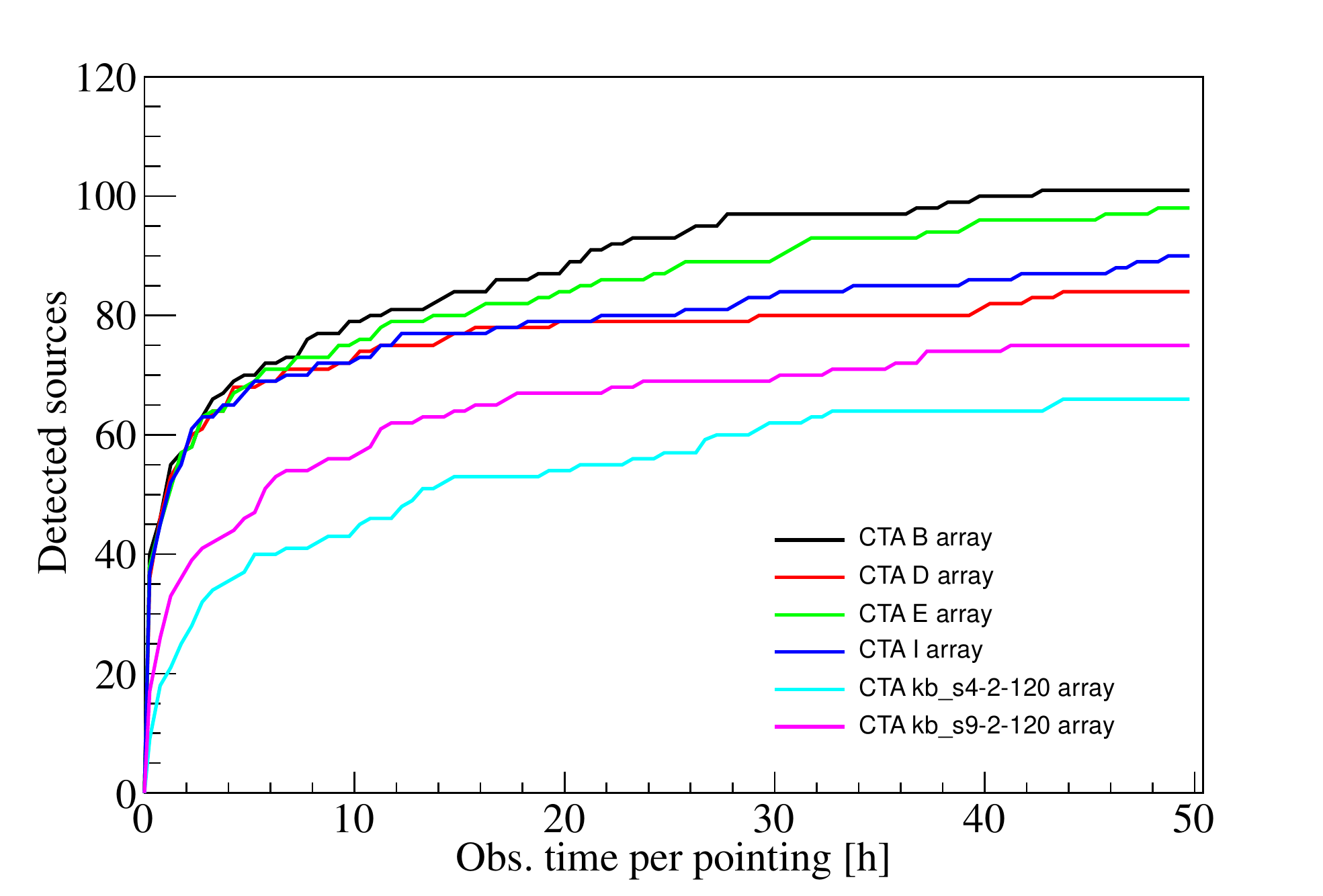}
\hfil
\caption{Cumulative distribution of the number of {\em Fermi}/LAT {\em Galactic} sources detected by CTA at the 5$\sigma$ level as a function of observing time and for various array configurations.  The parent population consists of 196 sources in the 2FGL catalog within $|b|<2^{\circ}$.}
\label{nmfig1}
\end{figure}

{\em Galactic Surveys.} For Galactic sources, we consider all associated/unassociated sources at low Galactic latitude ($|b|<2^{\circ}$). The Galactic sample includes
high-mass binaries, supernova remnants (SNRs), pulsar wind nebulae (PWNe) and unassociated sources. Throughout, we have excluded sources listed in the Second {\it Fermi}/LAT AGN Catalog \citep{2011arXiv1108.1420T} and highly variable unidentified sources (with a {\it Fermi}/LAT Variability Index (VI) greater than 41.6, as described in the catalog) in order to strictly collect {\it bona fide} or potential Galactic sources. This strategy  resulted in a total of 196 tentatively tagged sources in the Galactic category at $|b|<2^{\circ}$. Repeating the exercise for sources at $|b|<5^{\circ}$ increases the initial sample to 297.  Given the limited spatial information from the 2FGL, we model the entire sample as point sources.   The key variable to consider is the time employed  per pointing. Figure~\ref{nmfig1} shows that $\geq 70$ 2FGL sources (or 35\% of the initial sample) are detected by CTA with exposure times of 5 hr or more when using full-array configurations (B, D, E or I). The performance of smaller subsets of the array (s4-2-120 and s9-2-120 with 4 and 9 medium-sized telescopes, respectively) has also been considered. While not as effective as a fully dedicated array, the fraction of detected sources remains significant (Fig.~\ref{nmfig1}). One option would be to use large-sized telescopes for extragalactic sources (which tend to have soft spectra) and small/medium-sized telescopes for the Galactic Plane sources (which tend to have hard spectra). Although this is not possible within a survey strategy, note that up to $\approx$ 50\% of the 2FGL sample of Galactic sources are within the reach of CTA, using exposure times as long as 50 hr/source (Tab.~\ref{comp1}). The best array configurations for this are B and E. With the observed {\em Fermi}/LAT source density in the Galactic plane ($|b| < 5^{\circ}$) and a 25 square degree FoV gives an average of 2 sources per field, a total of 4000 hr would be needed to complete a targeted survey returning 50\% of the {\em Fermi}/LAT Galactic catalog.

\begin{figure}[t]
\hfil
\includegraphics[width=\linewidth]{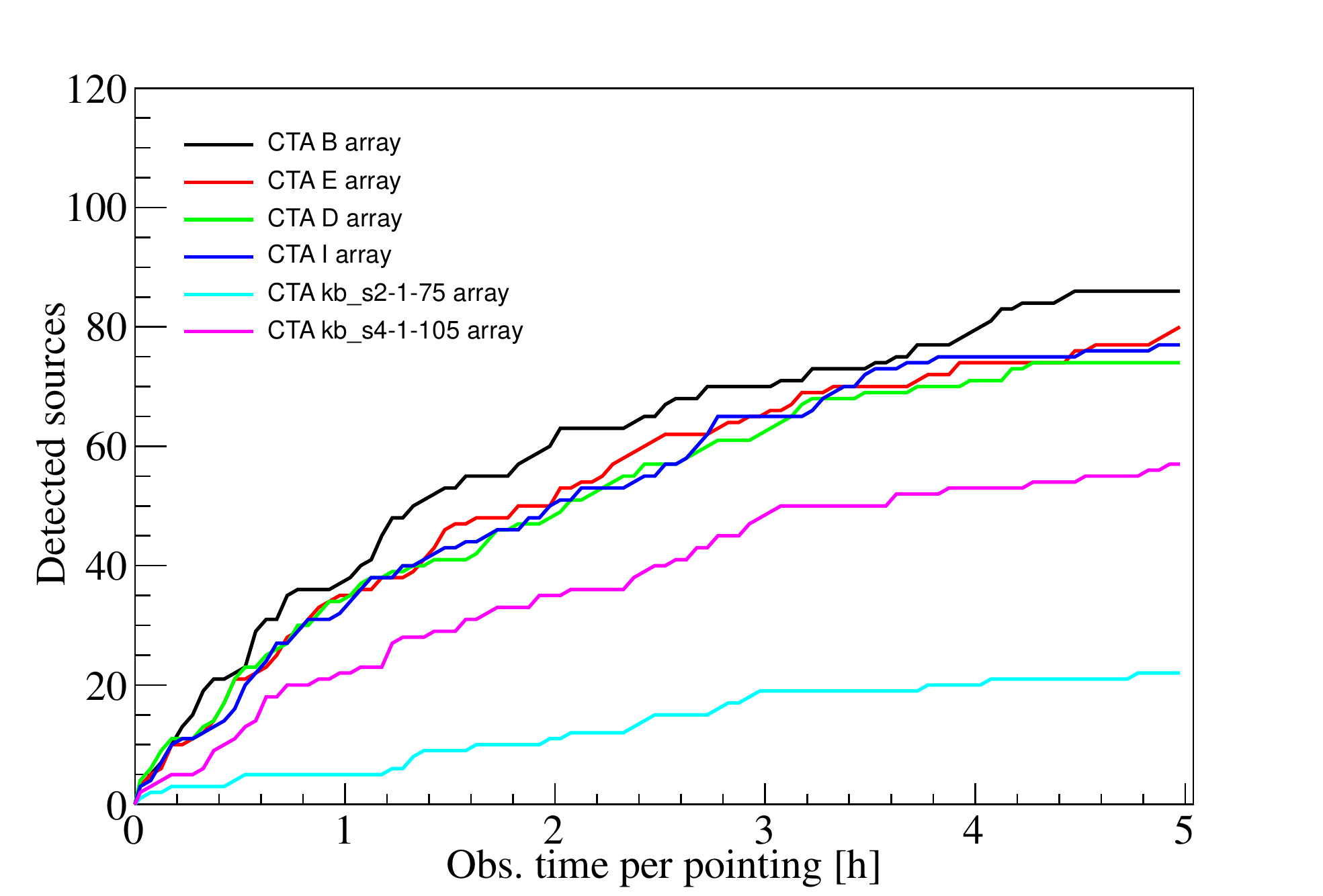}
\hfil
\caption{Cumulative distribution of the number of {\em Fermi}/LAT {\em extragalactic} sources detected by CTA at the 5$\sigma$ level as a function of observing time and for various array configurations.  The parent population consists of 561 sources in the 2FGL catalog identified as extragalactic.}
\label{nmfig2}
\end{figure}

{\em Extragalactic Surveys.} In the case of extragalactic sources, we consider a subset of 561 {\em Fermi}-labelled extragalactic sources \citep{cta_agn}. For the latter, the extrapolated VHE spectra were attenuated using current estimations of the EBL absorption as a function of redshift \citep{EBL_Franceschini}.  For nearby hard sources $\Gamma<2$, a straight extrapolation could create runaway integrations, therefore we applied an {\em ad hoc} broken power law with $\Gamma$ = 2.5 starting at 100 GeV to soften such spectra. Using the expected effective areas and background rates from Monte Carlo simulations, our models find that a CTA all-sky survey with typical exposure time of 0.5 hr (as envisioned in \S\ref{strategy}) would detect only $\approx$ 20\ {\em Fermi}/LAT extragalactic sources over the whole sky (Fig. ~\ref{nmfig2}), {\em i.e.} 5 within the 1/4th of the sky observable with good zenith angle (assuming the sources are distributed uniformly over the sky). With 5 hr/pointing the number of sources increases to 80 ({\em i.e} 20 in practice). The total number of {\em Fermi}/LAT extragalactic sources detectable with CTA exposure times up to 50 hr/source is $\geq 170$ (30\% of the initial sample) with the most favorable array configurations (B and E, Tab.~\ref{comp1}). 

\begin{table} 
\centering           
\begin{tabular}{lcc}
\hline
{Array}       & {Extragalactic}  & Galactic      \\
& (561 sources) & (196 sources)\\
 \hline
B & 192 & 101\\
D & 138 & 84\\
E & 171 & 98 \\
I & 159 & 90\\
\hline
\end{tabular}
\caption{Number of {\em Fermi}/LAT sources selected from the 2FGL catalog that are detected by CTA using exposure times up to 50 hr and various array configurations (see also Fig.~\ref{nmfig1}-\ref{nmfig2}).\label{comp1}}
\end{table}

\subsection{SNRs, PWNe in the Galactic Plane and source confusion\label{mr}}
The population study of PWNe and SNRs, the two main classes of VHE sources in the Galactic Plane, is based on the morphological and spectral characteristics of three representative shell-type SNRs (RX~J1713.7$-$3946, Vela~Jr and RCW~86) and PWNe (G21.5$-$0.9, Kes~75 and HESS~J1356$-$645) as measured with HESS \citep{c:renaud11,cta_snr}. Monte-Carlo simulations of these two source classes were carried out for different CTA array layouts, assuming a uniform exposure time of 20 hr everywhere along the Galactic Plane (giving a sensitivity of about 2 mCrab for a point source with spectral index $\Gamma$=2.5). This is a higher exposure time than in the initial first-year survey that is detailed later (with a uniform sensitivity corresponding to 8 hr of exposure time, \S\ref{strategy}) but a reasonable expectation of the exposure CTA can ultimately achieve after several years of operation.

For SNRs, about 20 to 70 SNRe are detected by CTA (configurations I and D, optimized for providing the best sensitivity over the whole energy range or above 1 TeV, respectively). VHE morphology is a powerful discriminant to identify shell-type SNRs but only a small fraction (7 to 15 sources) will be resolved {\em i.e.} those for which a  shell-type fit on the source radial profile is favored at $>$ 3$\sigma$ over a simple gaussian fit. The above-mentioned numbers are obtained by assuming a Galactic core-collapse SN rate of 2.5 century$^{-1}$ \citep[see][]{c:li11} and a timescale during which a SNR shines in the TeV domain of 5 kyr.

For PWNe, 300 to 600 PWNe should be detectable with the I or D configurations, assuming that the lifetime of TeV-emitting leptons in such sources (with a Galactic rate of 2  per century) amounts to $\sim$40 kyr ({\em i.e.} equal to the radiative timescale in a 3 $\mu$G magnetic field, as estimated in several PWNe with HESS such as Vela X \citep{c:dd09}).

To evaluate source confusion, the Galactic source distribution model of \citep{c:renaud11} was used to estimate the fraction of sources per square degree along each line-of-sight within -60$^{\circ}$ $<$ $\ell$ $<$ 60$^{\circ}$ and $|b|$ $<$ 5$^{\circ}$ (see also \S\ref{sim}). At first order ({\em i.e.~}neglecting the local variations at the locii of the spiral arm tangents), the resulting Galactic distribution is well fit with a 2D Gaussian lying at the Galactic center position, with a standard deviation $\approx$ 40$^{\circ}$ and 0.5$^{\circ}$ in $\ell$ and $b$, respectively and a maximal value of $\sim$ 4 (N$_{{\rm PWN}}$/500) sources per square degree. This implies that CTA should detect almost 200 sources in the central regions of the Galaxy, at $|\ell|$ $<$ 30$^{\circ}$ and $|b|$ $<$ 0.5$^{\circ}$, {\em i.e.~}$\sim$ 3 sources per square degree on average. Given that a large fraction of VHE-emitting (middle-aged) PWNe are expected to be extended (on scales of $\sigma$ $\sim$ 10--30 pc = 0.1$^{\circ}$--0.3$^{\circ}$ at 6 kpc), source confusion within the Galactic Plane survey performed with CTA will be an issue. Possible mitigating strategies include source identification using the highest energies, where the angular resolution improves and PWNe are more compact.

\subsection{Simulated CTA Galactic Plane survey \label{sim}}
To illustrate the potential of a Galactic Plane survey, we simulated scanning observations using the ctools, following the strategy identified in \S\ref{strategy} (a row of 60 pointings of 4 hr in steps of 2$^\circ$ along $b=0^\circ$). The on-axis 68\% containment radius for array configuration E  is about 7$^\prime$ at 100 GeV, about 3$^\prime$ at 1 TeV, and about 2$^\prime$ above 10 TeV. Two different population models were used to bracket the anticipated content in VHE emitters.

The first (model I) is based on the VHE source population model presented in \citep{2008AIPC.1085..886F}. The expected VHE emission from SNRs  is derived from a prescription for hadronic interactions of freshly-accelerated cosmic rays and a Sedov law for the size evolution. Monte-Carlo sampling of the spatial and energy distribution of supernova produces realizations of a Galactic population of VHE SNRs. The global model has free parameters - supernova rate, SNR TeV lifetime, explosion energy conversion efficiency, average density, scale height - which were fitted so that the model population properties match the distributions of fluxes, positions, and angular sizes of observed VHE sources. The contribution of PWNe has been added to the VHE population model described in \citep{2008AIPC.1085..886F}. PWNe are modeled from the ATNF pulsar catalog\footnote{see \url{http://www.atnf.csiro.au/research/pulsar/psrcat} \citep{2005AJ....129.1993M}}. For each pulsar, the age is used to obtain a gamma-ray luminosity from a fit to the observed VHE gamma-ray luminosity versus age relation, and a size from a fit to the observed size versus age relation. Given that the oldest VHE PWNe known are $\sim 10^6$ yr old, a cutoff at this age is applied to the luminosity-age relation. With these hypotheses, the PWN model population has no free parameter. The SNR and PWNe populations are added and the free parameters for the SNR population is readjusted so that the resulting total population is consistent with the observed source properties, with approximately half of the HESS sources being explained by SNRs and the other half by PWNe. A typical realization of the total Galactic population has about 4 times more SNRs than PWNe. 

The second (model II) focuses on PWNe only. PWNe are assumed to be produced at a rate of $\approx 2$ per century and have a lifetime of 40 kyr. Their distribution in longitude-latitude-distance is taken from the Galacto-centric SNR distribution \citep{1998ApJ...504..761C} with the spiral arm pattern of \citep{2008AJ....135.1301V} and a scale height of 130 pc. Simulated source properties are sampled from the observed distributions: a normal distribution of the logarithm of the 1-10 TeV luminosity with a mean of 34.4 and a standard deviation of 0.6 $\log$(erg/s), a uniform distribution between 2.0 and 2.5 for the spectral index $\Gamma$, and a uniform distribution between 5 and 30 pc for the size. About 500 PWNe are simulated based on the above assumptions within $|\ell|=60^\circ$ and $|b|=5^\circ$. The dominant source populations in the Galactic Plane are, at present, PWNe and SNRs but additional sources (binaries, dark accelerators, stellar clusters, diffuse emission etc.) are/could be expected. We have not attempted to take these into account in model I or II because of the very large uncertainties in the number and VHE properties of such sources. A better characterization of these sources would be a major goal of the survey.

\begin{figure*}
\centering
\resizebox{\hsize}{!}{\includegraphics{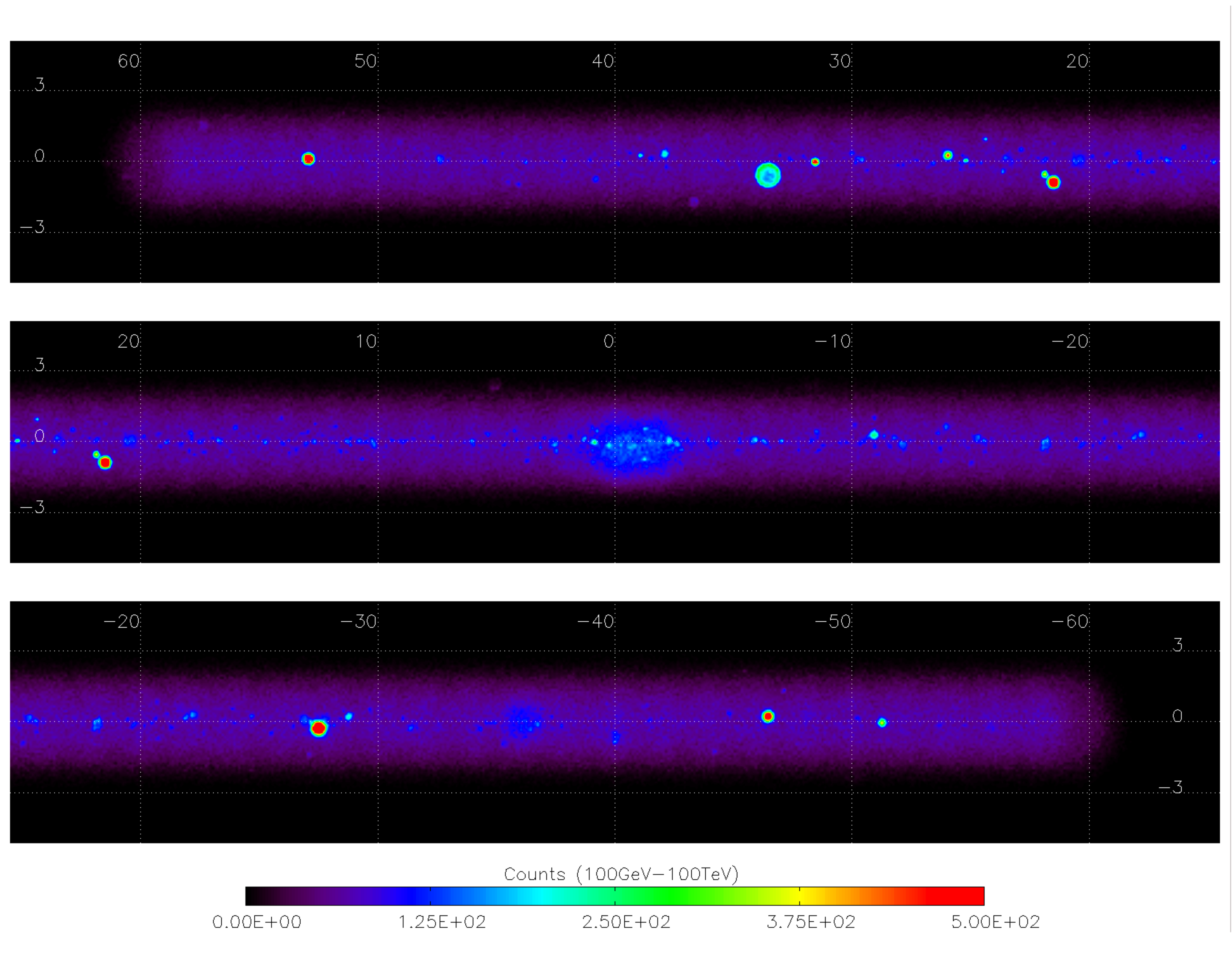}}
\caption{Simulated image of a 240 hr long CTA Galactic Plane survey using population model I as input.}
\label{cta_gps1}
\end{figure*}

\begin{figure*}
\centering
\resizebox{\hsize}{!}{\includegraphics{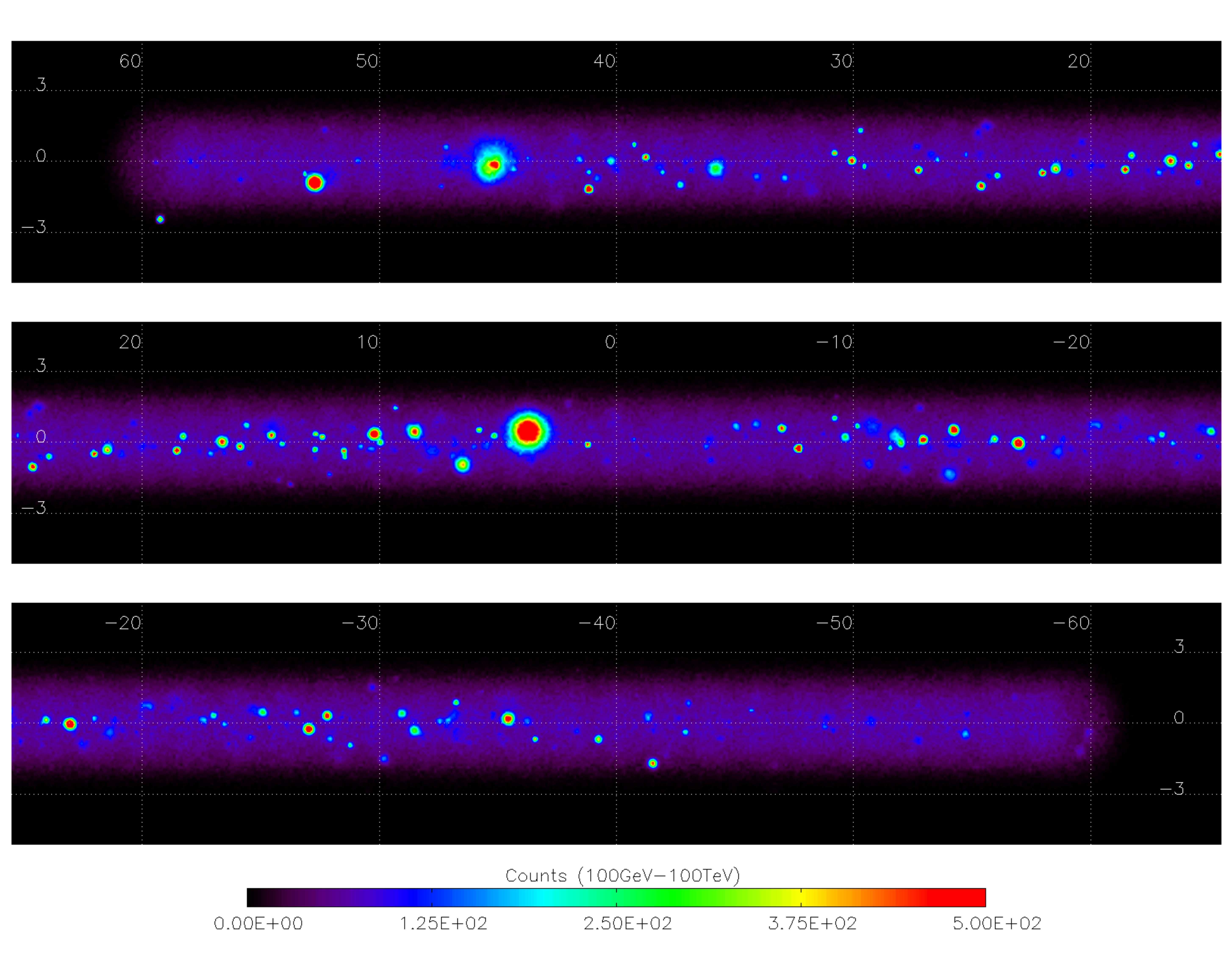}}
\caption{Simulated image of a 240 hr long CTA Galactic Plane survey using population model II as input.}
\label{cta_gps2}
\end{figure*}

Simulated counts maps of the full Galactic Plane survey for the two population model are shown in Figs.~\ref{cta_gps1}-\ref{cta_gps2}. The intensity distribution of a SNR is assumed to be a projected shell, the shell having a thickness equal to 10\% of its size. A 2D Gaussian intensity distribution is adopted for PWNe. The two input populations result in different VHE skies, illustrating the potential of CTA for population studies. Model II has many more bright and extended objects. Model I is dominated by a handful of bright PWNe, some of them being quite extended ({\em e.g.} close to the Galactic Center) ; the rest is then composed of numerous fainter SNR clustered along the plane. Both cases suggest that source confusion will likely be a challenge within the Galactic Plane (see above, \S\ref{mr}). 


\section{Discussion}
We have investigated two survey programs benefiting from the increased sensitivity and field-of-view that are planned for CTA. Realistic simulations using the CTA responses show (\S3) that uniform coverage of the Galactic Plane within $|\ell |=60^\circ$ and $|b|=2^\circ$ can be achieved down to a 3 mCrab sensitivity using 250 hr of observing time (Galactic Plane survey, \S\ref{gps}). A wide survey covering 1/4th the sky is also possible down to the 20 mCrab limit using 370 hr of observing time (all-sky survey, \S\ref{allsky}). These assume a sequential observing strategy where the surveyed  area is sampled with a high-sensitivity, relatively narrow field. A promising alternative is to use the divergent observing mode where the surveyed area is sampled with a low-sensitivity, wide field (\S\ref{divergent}). Preliminary studies show the all-sky survey could then be achieved down to the same sensitivity using only 100 hr of observing time. 

Both surveys can be achieved within the available observing time in a year of operations but may need to be spread over several years for visibility constraints (these were not taken into account here so the exact scheduling and accessible sky area for each CTA site remain to be determined). The sensitivities that are reached are competitive in the multi-wavelength context (\S2), complementing well the surveys carried out by the {\em Fermi}/LAT (at lower energies) and EAS (at higher energies). In terms of implementation, there may be some practical advantage in using subarrays with large-sized telescopes concentrating on the extragalactic sky while the others work on Galactic projects. The loss in sensitivity (and angular resolution, \S\ref{sub}) is about a factor 2 when using a subarray with 4 large-sized telescopes or a subarray with 9 medium-sized telescopes (Fig. \ref{subarray}). For end-to-end spectral coverage and high angular resolution, the array must be fully dedicated.

The Galactic Plane survey is expected to lead to the detection of $\geq 70$ VHE counterparts to sources listed in the second {\em Fermi}/LAT catalog (\S\ref{fermi}). The expected number goes down to $\geq 50$ with a 9 telescope subarray (Fig.~\ref{nmfig1}). Population models for PWNe and SNRs based on current knowledge predict the detection of hundreds of sources with such a survey, source confusion likely becoming a difficulty (\S\ref{mr}). Inversely, a uniform survey will allow unprecedented constraints on population models (\S\ref{sim}). A Galactic Plane survey helps pinpoint sources for detailed morphological, spectral or timing studies, which will be necessarily limited for faint sources. Some timing information may be available depending upon the survey implementation: with 60 pointings of 4 hr, each location is visited at least 8 times by runs of 0.5 hr, giving in principle a sensitivity to variability with amplitudes $\geq 20$ mCrab (for instance for gamma-ray binaries \citep{cta_binaries}). 

A small number of detections are expected in the all-sky survey, based on current knowledge. The cost in observing time is reasonable if the survey can be achieved  within 100 hr in divergent mode. Divergent pointings would also allow to probe for variable sources with amplitudes $60-100$ mCrab, an attractive feature given that blazars are known to be variable and constitute the most numerous population of VHE sources in the extragalactic sky. The price to pay is the specific analysis tools that would need to be developed. A handful of counterparts to {\em Fermi}/LAT sources should be seen by a wide/shallow survey (\S\ref{fermi}). Predictions based on blazar population modeling (the dominant extragalactic source population) point to 10-20 detections (\S\ref{blazar}). Note though that these estimates do not take into account blazar flaring activity. For instance, the study presented in \S \ref{blazar} assumes an average blazar spectral energy distribution. Incorporating the poorly known duty cycle of blazars in the luminosity function models remains very difficult and could affect the numbers significantly.

Another approach to extragalactic population studies is to target known candidate sources, albeit at the cost of observational bias. Selecting times of flaring, as determined from other wavelengths, also increases the chance for AGN detections. Such strategies have enabled current IACTs to detect  $\approx$ 50 extragalactic sources. A targeted survey would still cover a significant area of the sky, allowing for serendipitous discoveries of sources in the FoV of the target ({\em e.g.} IC 310 in the FoV of NGC 1275 \cite{MAGIC_IC310}). For instance, the $\approx$ 20 brightest counterparts of {\em Fermi}/LAT sources observable by CTA are detected with a 5 hr exposure on each source, requiring a reasonable total time of 100 hr to complete (\S\ref{fermi}). The survey would cover a total area $\approx$ 200-400 square degrees (0.5-1\% of the sky), depending on the fraction of the FoV covered with uniform sensitivity (10-20 square degrees). Using only large-sized telescopes would make sense for targets affected by extragalactic background light absorption, although their smaller FoV reduces the survey footprint, freeing the rest of the array for the Galactic Plane survey. The area covered is much smaller than the envisioned ``all-sky survey" but with much better sensitivity (typically 3 mCrab compared to 20 mCrab, \S\ref{strategy}). At this sensitivity, a couple of serendipitous blazar detections are to be expected besides the targeted  {\em Fermi}/LAT sources (Tab.~\ref{tab:count_7}). As a follow-up, continuous regions of interest could be identified and imaged deeply ($\geq$ 5 hr) to complement the initial targeted survey. Such a explorative survey could aim for well-mapped areas at other wavelengths and be oriented to guide the design of subsequent observations. After several years, targeted observations at various extragalactic targets will add up to an increasingly wider portion of the sky surveyed, albeit not with uniform sensitivity, much like the coverage of the hard X-ray sky achieved by INTEGRAL/IBIS \citep{2010ApJS..186....1B}.

\section{Conclusion}
CTA will allow a survey of the inner Galactic Plane to unprecedented sensitivity ($\approx 3$ mCrab), close to the confusion limit, using $\approx 250$ hr of observing time. Simulations find hundreds of sources can be detected by the survey, enabling population studies and to pinpoint the most interesting sources for deeper follow-up. A Galactic Plane survey should be a major objective of CTA. A wide-area ``all-sky" survey down to 20 mCrab is also feasible using $\approx 400$ hr of observing time using standard techniques, or 100 hr using divergent pointing mode. Detailed studies of this mode, which takes advantage of the large number of telescopes in the CTA array, remain to be carried out. The prime motivation for such a survey is the search for new, unsuspected classes of VHE-bright sources (extragalactic ``dark accelerators") --- admittedly a gamble, but one with large payoff.

\section*{Acknowledgements}
GD thanks D. Horan \& S. Wakely for help with TeVCat (Fig.~1). GD and PM acknowledge support from the European Community via contract ERC-StG-200911. NM, TH and JLC acknowledge the support of the Spanish MICINN under project code FPA2010-22056-C06-06.  NM gratefully acknowledges support from the Spanish MICINN through a Ram\'on y Cajal fellowship. The research leading to these results has received funding from the
European Union's Seventh Framework Programme ([FP7/2007-2013]
[FP7/2007-2011]) under grant agreement n$^\circ$262053.
We also gratefully acknowledge support from the following agencies and organisations:
Ministerio de Ciencia, Tecnolog\'ia e Innovaci\'on Productiva (MinCyT),
Comisi\'on Nacional de Energ\'ia At\'omica (CNEA) and Consejo Nacional  de
Investigaciones Cient\'ificas y T\'ecnicas (CONICET) Argentina; State Committee
of Science of Armenia; Ministry for Research, CNRS-INSU and CNRS-IN2P3,
Irfu-CEA, ANR, France; Max Planck Society, BMBF, DESY, Helmholtz Association,
Germany; MIUR, Italy; Netherlands Research School for Astronomy (NOVA),
Netherlands Organization for Scientific Research (NWO); Ministry of Science and
Higher Education and the National Centre for Research and Development, Poland;
MICINN support through the National R+D+I, CDTI funding plans and the CPAN and
MultiDark Consolider-Ingenio 2010 programme, Spain; Swedish Research Council,
Royal Swedish Academy of Sciences financed, Sweden; Swiss National Science
Foundation (SNSF), Switzerland; Leverhulme Trust, Royal Society, Science and
Technologies Facilities Council, Durham University, UK; National Science
Foundation, Department of Energy, Argonne National Laboratory, University of
California, University of Chicago, Iowa State University, Institute for Nuclear
and Particle Astrophysics (INPAC-MRPI program), Washington University McDonnell 
Center for the Space Sciences, USA.





\bibliographystyle{elsarticle-num}
\bibliography{survey}







\end{document}